\documentclass[aps,prc,preprint,amsmath,amssymb,showpacs,preprintnumbers,superscriptaddress]{revtex4-1}
\usepackage{CJK}
\usepackage{graphicx}% Include figure files
\usepackage{dcolumn}% Align table columns on decimal point
\usepackage{bm}% bold math
\usepackage{color}
\usepackage{hyperref}% add hypertext capabilities
\allowdisplaybreaks

\begin{document}


\title{Intrinsic generation of angular momenta and entanglement in fission}

\author{B. Li}
\affiliation{State Key Laboratory of Nuclear Physics and Technology, School of Physics, Peking University, Beijing 100871, China}
\author{D. D. Zhang}
\affiliation{Institute of Theoretical Physics, Chinese Academy of Science, Beijing 100871, China}
\author{D. Vretenar}
\email{vretenar@phy.hr}
\affiliation{State Key Laboratory of Nuclear Physics and Technology, School of Physics, Peking University, Beijing 100871, China}
\affiliation{Physics Department, Faculty of Science, University of Zagreb, 10000 Zagreb, Croatia}
\author{T. Nik\v si\' c}
\affiliation{Physics Department, Faculty of Science, University of Zagreb, 10000 Zagreb, Croatia}
\author{P. W. Zhao }
\email{pwzhao@pku.edu.cn}
\affiliation{State Key Laboratory of Nuclear Physics and Technology, School of Physics, Peking University, Beijing 100871, China}
\author{J. Meng }
\email{mengj@pku.edu.cn}
\affiliation{State Key Laboratory of Nuclear Physics and Technology, School of Physics, Peking University, Beijing 100871, China}

\date{\today}

\begin{abstract}
Nuclear time-dependent density functional theory is used to investigate spin generation and entanglement of fission fragments in spontaneous fission of $^{252}$Cf, incorporating both axial and non-axial deformations.
Axially symmetric fission trajectories enforce strict constraints: counter rotation (twisting mode) along the fission axis and equiprobable bending/wriggling modes perpendicular to it. Non-axial modes broaden the distributions of fission fragment spin projection on the fission axis, and allow for axial (tilting) collective rotations, which are forbidden on axially symmetric trajectories. Mutual information analysis reveals that axial-symmetry breaking reduces spin-spin correlations along the fission axis of symmetric cases, while perpendicular correlations remain more resilient. The effect of triaxial degrees of freedom on the opening angle distribution between the spins of the fission fragments is analyzed.
\end{abstract}

\maketitle

Nuclear fission is a complex process in which a heavy nucleus splits into two or more fragments, releasing a substantial amount of energy along with neutrons, gamma rays, and other particles. One of the most intriguing and least understood aspects of fission is the generation of spins in the resulting fragments. The intrinsic spins and orbital angular momenta of these fragments play a crucial role in their de-excitation, influencing gamma-ray emission and neutron evaporation~\cite{Schmidt2018RPP,Bender2020,Schunck22PPNP}. Even though theoretical \cite{Nix1965NP,Rasmussen1969NPA,ZielinskaPfabe1974PLB,Bonneau2007PRC}, and experimental \cite{Wilhelmy1972PRC,Wolf1976PRC} investigations started decades ago, the mechanisms governing the origin and distribution of these spins remain a subject of active debate.

Recent experimental advances~\cite{Wilson2021Nature,Marin2021PRC,Marin2022PRC,Giha2023PRC,Marin2024PRC}, including high-resolution gamma-ray spectroscopy and correlated fragment-neutron-gamma measurements, have provided new insights into fragment spin polarization and alignment. These observations suggest that spins arise from a combination of collective dynamics, quantum correlations at scission, and nuclear shell effects. Furthermore, the possibility of quantum entanglement between the fragments -- stemming from their shared origin in a highly correlated nuclear system -- opens a new frontier in the study of fission dynamics.

In Ref.~\cite{Wilson2021Nature}, extensive experimental data on fragment spins were analyzed to test competing theoretical models. The study found no significant correlation between the spins of complementary fragments, along with a strong mass-dependent variation in spin magnitudes, exhibiting a distinctive sawtooth pattern. Crucially, the fragment spins appeared independent of their partners' mass or charge, leading the authors to conclude that spin generation occurs post-scission in an uncorrelated manner. These findings have spurred a wave of theoretical investigations using both microscopic~\cite{Bulgac2021PRL,Bulgac2022PRL,Scamps2022PRC,Scamps2023PRC,Marevif2021PRC,Marevic2025} and phenomenological approaches~\cite{Randrup2021PRL,Randrup2022PRC,Randrup2022PRC2,Randrup2023PRC,Zhou2024,Dossing2024PRC,Shneidman2025PRC}.

Contrary to the post-scission interpretation, studies based on time-dependent density functional theory (TDDFT)~\cite{Bulgac2021PRL,Bulgac2022PRL,Scamps2022PRC,Scamps2023PRC} indicate that fragment spins are determined before scission, exhibiting strong correlations dominated by collective bending and twisting modes. These spins are primarily oriented perpendicular to the fission axis~\cite{Scamps2022PRC,Scamps2023PRC2}, with their magnitudes heavily influenced by shell structure and deformation~\cite{Bertsch2019PRC,Randrup2021PRL,Marevif2021PRC,Dossing2024PRC}. The opening angle distribution between the spins of fission fragments has been investigated using several theoretical approaches \cite{Randrup2022PRC,Randrup2022PRC2,Bulgac2022PRC,Scamps2024PRC}.
 While Wilson et al.~\cite{Wilson2021Nature} argued for post-scission spin generation, multiple studies~\cite{Bulgac2022PRL,Randrup2021PRL,Scamps2023PRC2} demonstrate that statistical and dynamical effects can decouple spins even before scission. Additionally, neutron and gamma emissions~\cite{Stetcu2021PRL} may significantly distort initial spin correlations, masking the true primordial distributions.

The observed sawtooth pattern in spin magnitudes reproduced in both microscopic~\cite{Marevif2021PRC,Marevic2025} and phenomenological~\cite{Randrup2021PRL,Dossing2024PRC} models is linked to shell closures and fragment deformations. Intriguingly, this pattern may only emerge after neutron and gamma emissions~\cite{Stetcu2021PRL,Marevif2021PRC}, implying that experimentally measured spins do not directly reflect scission dynamics. Complementary techniques, such as photon angular correlations~\cite{Randrup2022PRC} and $E2$ transition analyses, offer promising avenues to probe fragment spin orientations. Meanwhile, studies incorporating octupole deformations~\cite{Scamps2023PRC2} and Coulomb torque effects~\cite{Scamps2022PRC,Randrup2023PRC} reveal additional complexities in spin generation mechanisms.

These advances underscore the intricate nature of fragment spin generation, with microscopic theories providing deeper insights while also highlighting discrepancies with experimental interpretations. However, one key limitation persists in most theoretical treatments: the assumption of axial symmetry in the fissioning system. In reality, non-axial deformations play an important role in both nuclear structure and dynamics~\cite{Bohr1975,RingManybody,Meng2016}, giving rise to phenomena such as wobbling \cite{Bohr1975} and chiral rotation~\cite{Frauendorf1997NPA}. Including triaxial degrees of freedom has already been shown to improve predictions for fission barriers in actinides \cite{Lu2012PRC,Lu2014PRC,Ryssens2023EPJA}.

This Letter presents the first investigation of the influence of non-axial trajectories on fission fragment (FF) spins and their correlations, carried out using time-dependent relativistic density functional theory. Pairing correlations are described within the Bardeen-Cooper-Schrieffer (BCS) approximation~\cite{Ebata2010PRC,Scamps2013PRC}. Unlike models based on the fully dynamical Hartree-Fock-Bogoliubov (HFB) theory of nuclear pairing~\cite{Bulgac2021PRL,Bulgac2022PRL,Scamps2022PRC,Scamps2023PRC}, the BCS approach does not fully account for dynamical effects of the pairing field~\cite{Magierski2018APP}. In the present study, large-amplitude collective motion is driven by the evolution of the nucleon density, while the pairing gap adapts at each time step to the instantaneous density. The BCS approximation greatly reduces the computational complexity of a full time-dependent HFB implementation, thereby making calculations feasible that break the axial symmetry of the fissioning nucleus. However, it is well known that dynamical models based on BCS can lead to violations of the local continuity equation \cite{Scamps2012PRC} and exhibit limitations in describing a superfluid fermionic gas \cite{Bulgac2007PRA}. Our approach builds on established fission modeling techniques \cite{Ren2020PRC,Ren2020PLB,Ren2022PRL,Ren2022a,Ren2022b,Zhang2024PRCa,Zhang2024PRCb,Li2023PRC_FTTD,Li2024PRCa,Li2024PRCb,Li2024PLB,Zhang2025PLB} (see Supplement~\cite{Li2025Supp} for details), focusing on spontaneous fission (SF) of $^{252}$Cf. This system is particularly instructive as its initial state carries zero total spin, allowing us to isolate how symmetry-breaking dynamics generates FF spins and entanglement between fragments.

In the left panel of Fig.~\ref{fig_ES}, we display the deformation energy surface for $^{252}$Cf, as function of the axially symmetric quadrupole $\beta_{20}$ and octupole $\beta_{30}$ deformation parameters (see Supplement~\cite{Li2025Supp} for the definition of $\beta_{\lambda\mu}$), obtained by self-consistent deformation-constrained relativistic density functional theory calculations in a three dimensional coordinate lattice space~\cite{Ren2017PRC,Ren2019SCPMA,Ren2020NPA,Li2020PRC,Xu2024PRC,Xu2024PLB,Xu2024PRL}, with the relativistic density functional PC-PK1~\cite{Zhao2010PRC} and a monopole pairing interaction.
The equilibrium minimum is at $(\beta_{20},\beta_{30}) = (0.3,0.0)$, and a single fission barrier is found at $(\beta_{20},\beta_{30}) \approx (0.6,0.0)$. The SF of $^{252}$Cf begins in the ground state potential well, traverses the fission barrier via quantum tunneling, and emerges with conserved energy on the post-barrier side. In the multidimensional deformation space, the probability for the fission path to reach any particular point on the post-barrier isoenergy surface is governed by the barrier characteristics and the collective inertia tensor.

%-----------------------------------------------------
\begin{figure}[!htbp]
\centering
\includegraphics[width=0.8\textwidth]{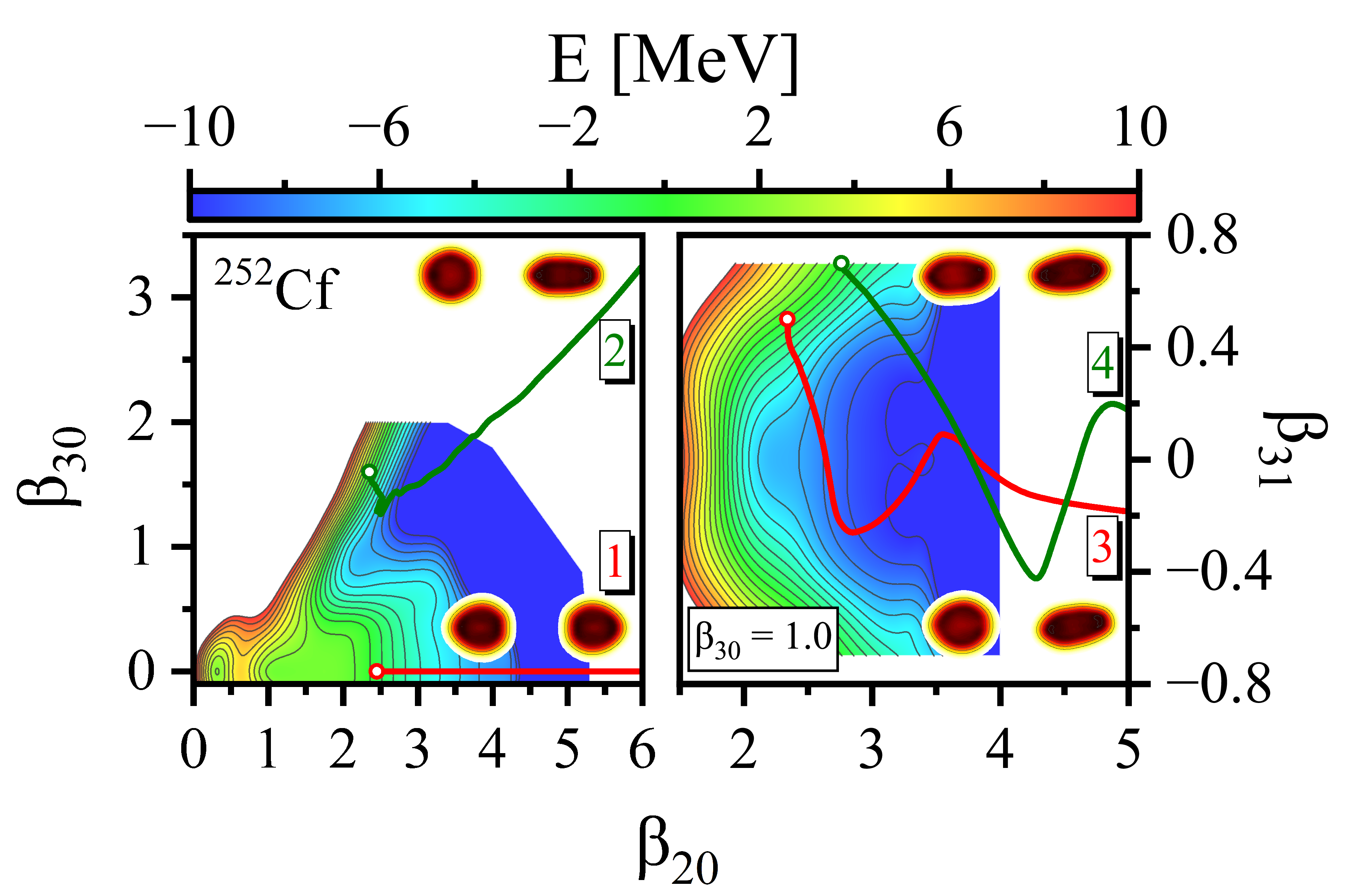}
\caption{(Color online) Left panel: Self-consistent deformation energy surface of $^{252}$Cf in the plane of axially symmetric quadrupole $\beta_{20}$ and octupole $\beta_{30}$ deformation parameters. Right panel: Same as in the left panel, but for
$\beta_{20}$ and the non-axial octupole ${\beta_{31}}$ deformation parameters, for a fixed value $\beta_{30} = 1.0$. Contours join points on the surface with the same energy, and the open dots correspond
to points on the iso-energy curve at the energy of the equilibrium minimum. The curves denote self-consistent TDDFT fission trajectories that start from the four initial points.
The corresponding density profiles after scission, when the distance between the two fragments equals $25$~fm, are shown in the insets.}
\label{fig_ES}
\end{figure}

While TDDFT provides an effective description of the evolution of independent nucleons in self-consistent mean-field potentials, this framework cannot describe quantum tunneling through classically forbidden regions of the collective deformation space \cite{Simenel2014PRC,Schunck2016PPNP,Bender2020}. Consequently, we initialize the TDDFT evolution of fission trajectories at exit points beyond the fission barrier (indicated by open circles in Fig.~\ref{fig_ES}), where the system emerges from the quantum tunneling regime. From these starting configurations, TDDFT simulates individual fission events by propagating nucleons independently through the scission point and into the post-scission dynamics. The trajectories 1 and 2 start from points that correspond to axially symmetric deformations, and this symmetry is preserved by the self-consistent evolution in time. Trajectory 1 is reflection symmetric (initial $\beta_{30}^{init} = 0$), while trajectory 2 starts from a point with finite octupole deformation (initial $\beta_{30}^{init} = 1.6$).

The right panel of Fig.~\ref{fig_ES} displays the deformation energy surface as a function of the quadrupole $\beta_{20}$ and the non-axial octupole coordinate $\beta_{31}$, for a fixed axial octupole deformation $\beta_{30}=1.0$. The surface exhibits remarkable softness along the $\beta_{31}$ direction, strongly motivating the inclusion of non-axial configurations. Accordingly, we examine two distinct non-axial trajectories (3 and 4) originating from post-barrier points with $\beta_{31}$ deformations of $0.5$ and $0.7$, respectively, while maintaining the energy of the equilibrium minimum. While we have also investigated configurations with finite $\beta_{32}$ deformations (see the supplemental material~\cite{Li2025Supp}), the increase in deformation energy is considerably steeper in $\beta_{32}$ than in $\beta_{31}$ direction. For $\beta_{31}$, we could select initial points with rather large deformation (non-axiality), while still at an energy close to the equilibrium minimum, that is, corresponding to spontaneous fission.

The insets of Fig.~\ref{fig_ES} show the corresponding FF density profiles after scission, when the distance between the centers of mass of the two fragments equals $25$~fm. For trajectory 1, both the FFs are nearly spherical. The heavy FFs are nearly spherical, and the light fragments are well-deformed for trajectories 2 and 3. For trajectory 4, both fragments are well-deformed. Due to breaking of axial symmetry for trajectories 3 and 4, the elongation axes of FFs do not align with the $z$-axis of the coordinate space lattice. When the fragments are separated by $25$~fm, the time evolution is stopped and spin distributions are evaluated by angular momentum projection (AMP). The time-evolution of one-body density profiles along the four trajectories are illustrated in the Supplement~\cite{Li2025Supp}.

\begin{figure}[!htbp]
\centering
\includegraphics[width=0.8\textwidth]{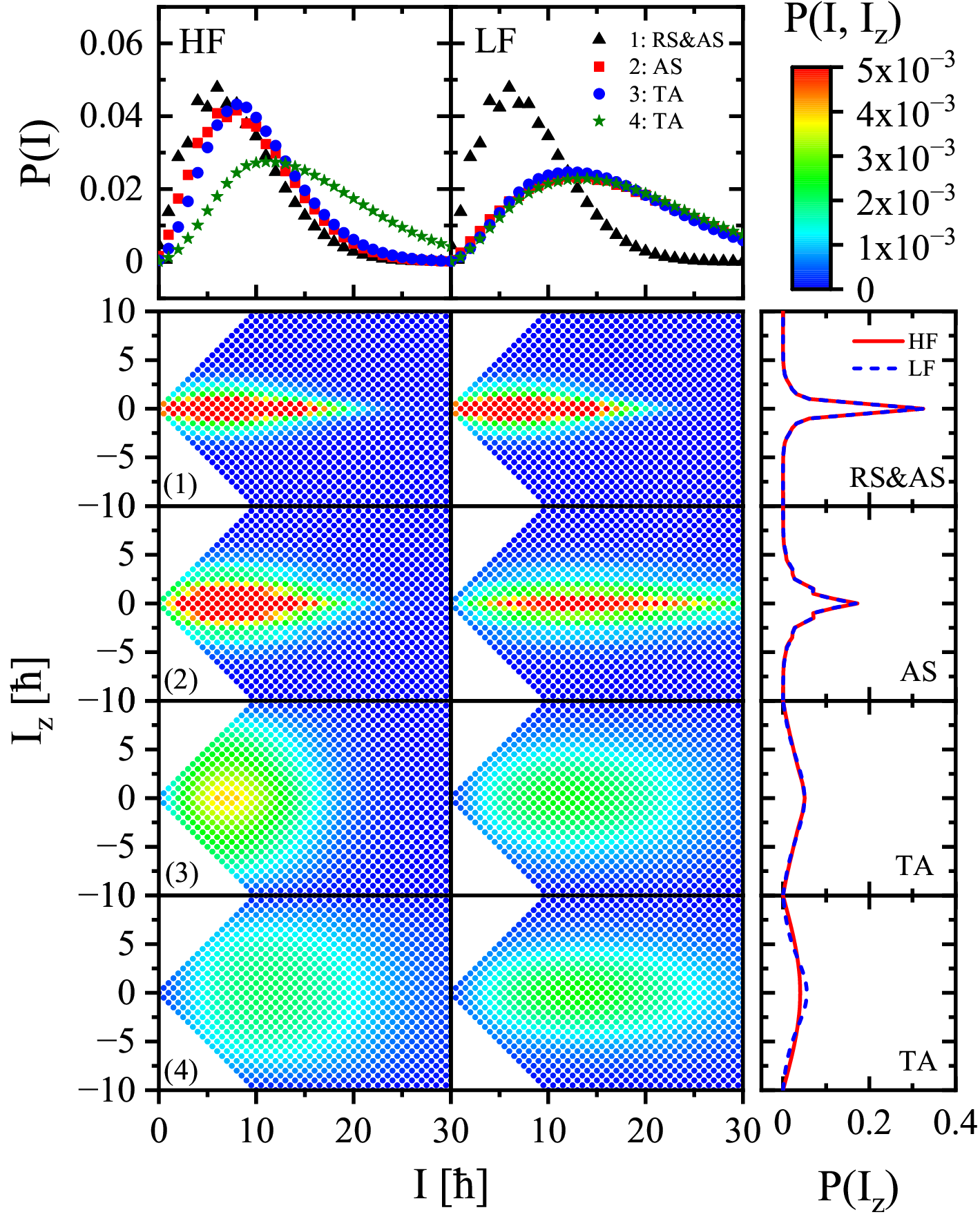}
\caption{(Color online) The first row shows the probability distributions of the spin $I$ for the heavy (left panel), and light FFs (right panel), obtained from the four TDDFT fission trajectories. In the second row, for trajectory 1, we plot the two dimensional probability distributions of the spin $I$ and its projection $I_z$ on the $z$-axis, for the heavy (left panel) and light FFs (middle panel). The corresponding marginalized distributions of $I_z$ are shown in the right panel. In the rows below, the corresponding distributions are plotted for trajectories 2, 3, and 4.
Trajectory 1 satisfies both reflection and axial symmetries (RS\&AS), while only  axial symmetry (RS) is conserved along trajectory 2. Two triaxial trajectories 3 and 4 (TA), break axial symmetry.}
\label{fig_AMP_fg}
\end{figure}

For each fragment, the probability distributions $P(I,I_z)$ of the spin $I$ and its projection $I_z$ on the $z$-axis, are
obtained by AMP \cite{KENJI1995IJMPE,Scamps2023PRC,Zhang2025PLB} (see Supplement~\cite{Li2025Supp} for details).
The distributions $P(I,I_z)$ for heavy (H) and light (L) FFs are shown in left and middle panels of Fig.~\ref{fig_AMP_fg}, respectively. Both integer and half-integer spins, and the corresponding projections are included. The distribution $P(I,I_z)$ is symmetric with respect to  $I_z=0$ for all trajectories, even when axial symmetry is broken.

Our calculations reveal systematic mass dependence in fragment spins, with the lighter fragment consistently exhibiting higher spin values (the first row of Fig.~\ref{fig_AMP_fg}) due to its greater deformation. This trend holds for all trajectories except the reflection-symmetric case (trajectory 1). For trajectory 1 (the second row, left and middle panels), the spin distribution $P(I,I_z)$ shows strong $I_z$-space localization, with probabilities falling below $10^{-3}$ for $|I_z|>2.5~\hbar$ and $I>23~\hbar$. Breaking reflection symmetry (trajectory 2) significantly enhances higher-$I_z$ probabilities for the heavy fragment, extending the probability $P > 10^{-3}$ region to $|I_z|= 4~\hbar$. In all cases, fragment spins align predominantly perpendicular to the fission axis. The right panels display the marginalized $I_z$-distributions $P(I_z)=\sum_{I} P(I,I_z)$. For both trajectories 1 and 2, axial symmetry and zero total angular momentum conservation enforce identical $P(I_z)$ distributions for light and heavy fragments. Notably, the $I_z = 0$ peak broadens when reflection symmetry is broken, indicating enhanced $I_z$ mixing.

The spin distributions $P(I,I_z)$ for non-axial trajectories 3 and 4, initialized with finite $\beta_{31}$ deformations, are presented in the fourth and fifth rows of Fig.~\ref{fig_AMP_fg}. While these distributions maintain similar spin magnitude ($I$) characteristics to their axial counterparts, they exhibit significantly greater dispersion in $I_z$-space. Importantly, in these non-axial cases, $I_z$ represents the spin projection along the computational lattice's $z$-axis rather than the fragment's symmetry axis. Consequently,
$I_z$ ceases to be a conserved quantum number for the total system, and FF spin correlations enforced by axial symmetry are no longer present. The right panels of Fig.~\ref{fig_AMP_fg} display the marginalized $P(I_z)$ distributions for trajectories 3 and 4, which show substantially broader profiles compared to axial cases. Furthermore, the relaxation of the axial symmetry constraint ($I_z^H+I_z^L=0$) permits asymmetric $I_z$-distributions between light and heavy FFs.

For the heavy FFs obtained from the trajectories 1, 2, and 3, the distributions $P(I)=\sum_{I_z} P(I,I_z)$ (shown in the first row of Fig.~\ref{fig_AMP_fg}) peak at $6~\hbar$, $8~\hbar$, and $8~\hbar$, respectively. The corresponding mass numbers are $126.0$, $139.5$ and $138.9$, and charge numbers are $49.0$, $53.6$, and $53.8$. The peak of $P(I)$ for the heavy FF in trajectory 4 (mass number $136.6$ and charge number $52.8$) is at $11~\hbar$, because this fragment is more deformed ($\beta_{20}\approx0.5$). The distributions $P(I)$ of the light FFs in trajectories 2, 3, and 4 are similar, and exhibit more dispersion than the symmetric trajectory 1.

The effect of the Coulomb force on spins of the FFs, as they move apart after scission, is illustrated for the two non-axial trajectories in Figs. 3 and 4 of the Supplement~\cite{Li2025Supp}.
It should be noted that the spin distribution could still evolve when the Coulomb interaction becomes negligible at large fragment separations, because two-body observables are not strictly conserved in the TDDFT framework.
Nevertheless, as we follow the fragment separation from 30 fm to 60 fm, this spurious effect on the spin distributions is rather small, as shown in Fig. 4 of the Supplement~\cite{Li2025Supp}.

\begin{figure}[!htbp]
\centering
\includegraphics[width=0.8\textwidth]{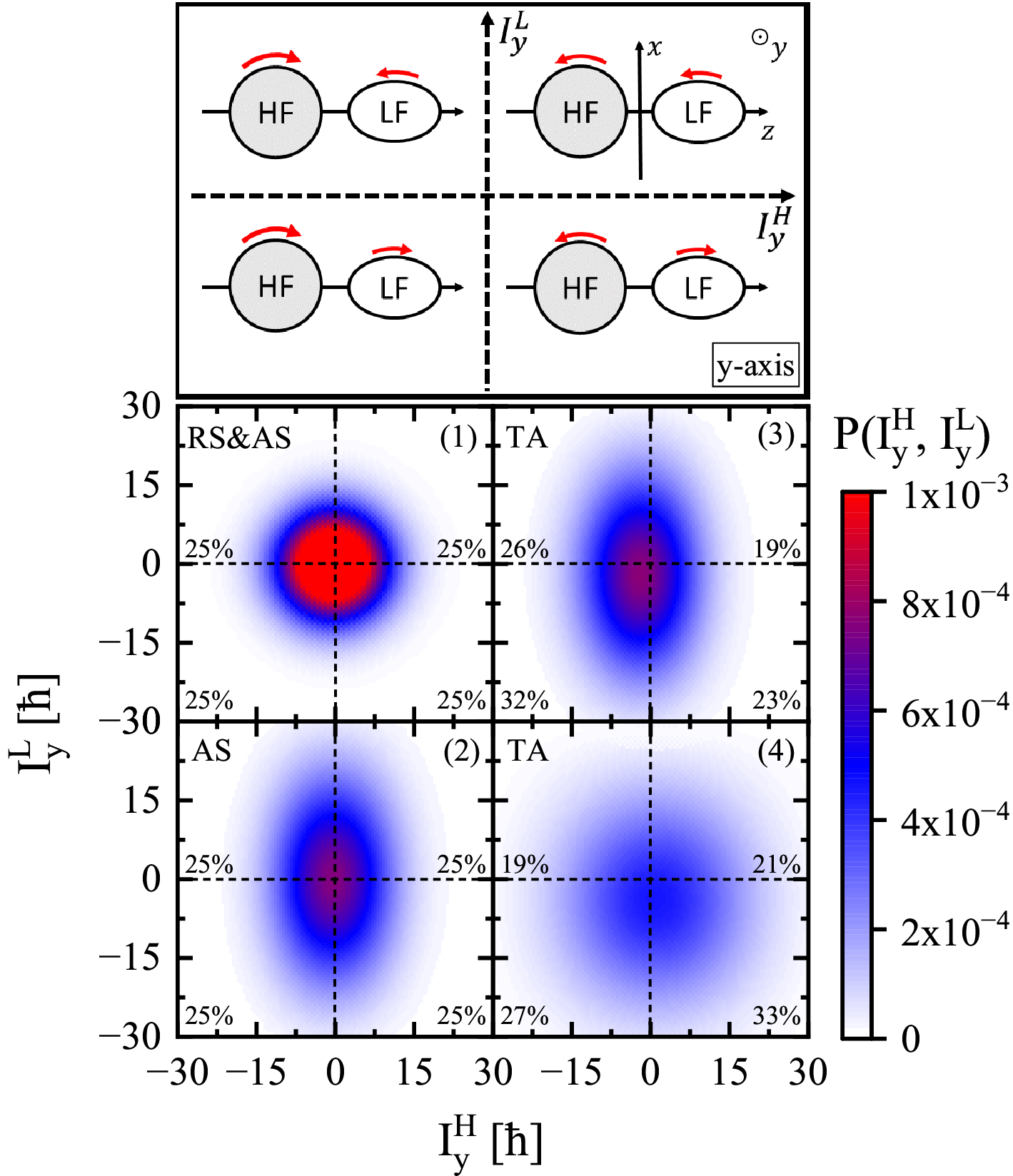}
\caption{(Color online) Panels (1)-(4): The joint probability distributions $P(I_y^H,I_y^L)$ for the projections of the FF spins along the $y$-axis, for fragments that are formed at the end of trajectories $1- 4$.
The percentages refer to the sum of probabilities in the corresponding quadrants. In the top panel the bending and wriggling modes of rotation are illustrated.}
\label{fig_AMP_y}
\end{figure}

To evaluate correlations between the intrinsic spins of the two FFs, we compute the joint distribution
$P(I_k^H,I_k^L)$, where $I_k^H$ and $I_k^L$ are components of the spin of the heavy and light FFs along the $k$-axis, respectively.
Since the deformation $\beta_{31}$ bends the nucleus in the $x$-$z$ plane (see Fig.2 in the Supplement~\cite{Li2025Supp}), the expectation values of the FFs' spins are predominantly along the $y$-axis. Thus, we first consider AMP along the $y$-axis.
Figure \ref{fig_AMP_y} presents the joint probability distributions $P(I_y^H,I_y^L)$ for the $y$-axis spin projections of complementary FFs generated on trajectories $1-4$. The $y$-axis projections (perpendicular to the reaction plane) reveal two fundamental rotational modes: 1) wriggling: correlated rotations ($I_y^H>0,~I_y^L>0$ and $I_y^H<0,~I_y^L<0$) and, 2) bending: anti-correlated rotations ($I_y^H>0,~I_y^L<0$ and $I_y^H<0,~I_y^L>0$) \cite{Nix1965NP}. For trajectory 1 (preserving both reflection and axial symmetries), the distribution forms a perfect circle in the $I_y^H$-$I_y^L$ plane. Breaking reflection symmetry (trajectory 2) distorts this distribution into an ellipse, with its major axis aligned along $I_y^L$ and minor axis along $I_y^H$ - a direct consequence of the light FF's larger spin components. Notably, both symmetric trajectories (1 and 2) show equal $25\%$ probabilities in each quadrant, indicating statistically equivalent likelihoods for bending and wriggling modes. This equiprobability reflects the underlying symmetry constraints governing these cases.

Breaking axial symmetry leads to asymmetric $P(I_y^H,I_y^L)$ distributions, manifesting as unequal probabilities for bending versus wriggling modes, as shown in the insets of Fig.~\ref{fig_AMP_y}. For trajectory 3 and the bending mode, we find $26\%$ probability for $I_y^H<0,~I_y^L>0$ and $23\%$ for $I_y^H>0,~I_y^L<0$.
For the wriggling mode, there is $19\%$ probability for $I_y^H>0,~I_y^L>0$ and $32\%$ for $I_y^H<0,~I_y^L<0$.
For the trajectory 4 the distribution $P(I_y^H,I_y^L)$ appears almost rotated by $\pi/2$.  In the case of bending, $19\%$ is the probability for $I_y^H<0,~I_y^L>0$ and $33\%$ for $I_y^H>0,~I_y^L<0$, while for the
wriggling mode $21\%$ is the probability for $I_y^H>0,~I_y^L>0$ and $27\%$ for $I_y^H<0,~I_y^L<0$.

\begin{figure}[!htbp]
\centering
\includegraphics[width=0.8\textwidth]{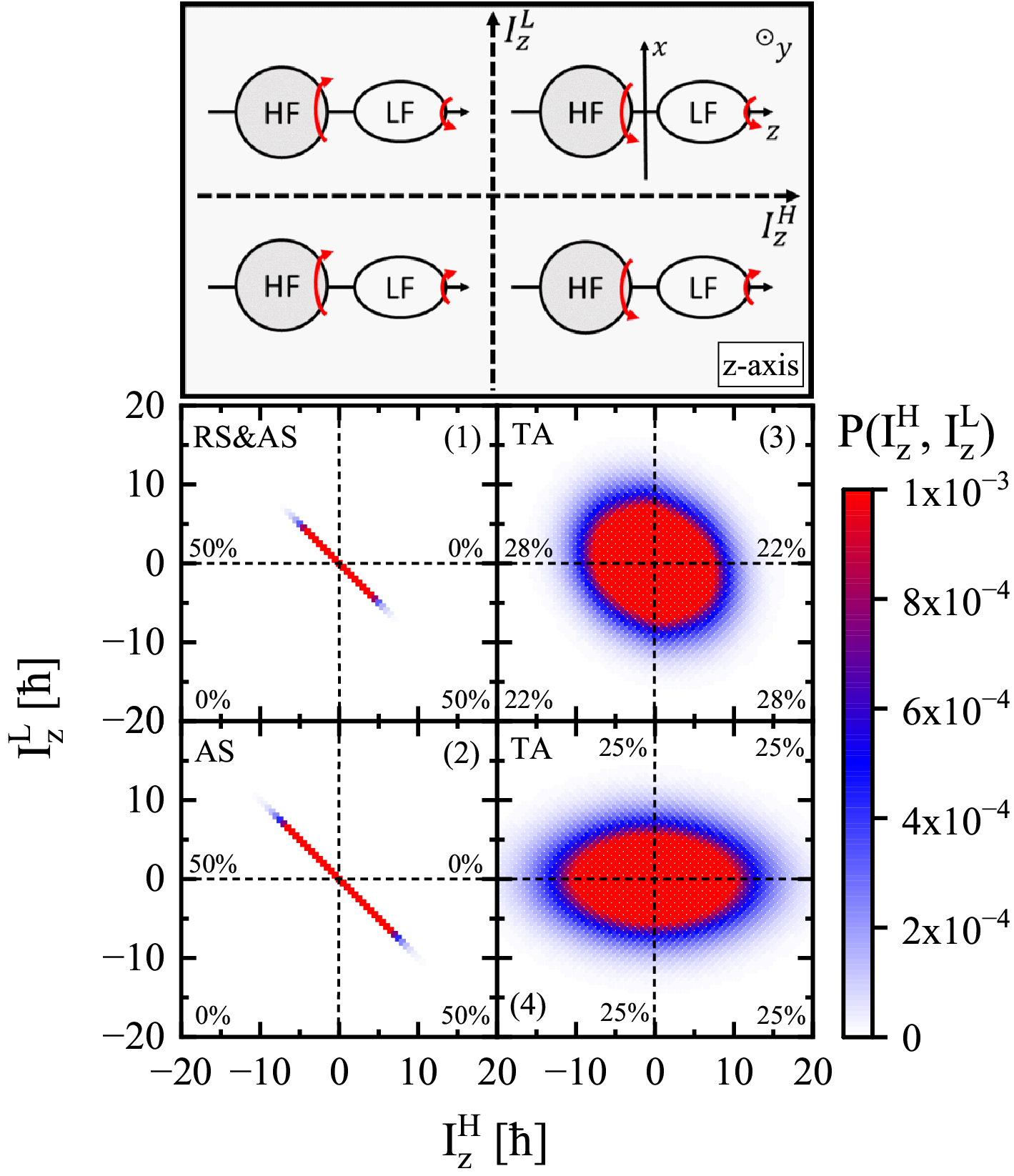}
\caption{(Color online) Same as Fig.~\ref{fig_AMP_y} but for projections along the $z$-axis. In the top panel the twisting and axial modes of rotation are illustrated.}
\label{fig_AMP_z}
\end{figure}

We now examine spin projections along the $z$-axis (the fission axis). This is the direction along which the FFs separate, and the corresponding joint probability distributions $P(I_z^H,I_z^L)$ for trajectories $1-4$ are shown in the panels (1)-(4) in Fig.~\ref{fig_AMP_z}, respectively. The rotational modes along this axis fall into two distinct categories based on the relative signs of $I_z^H$ and $I_z^L$.
For $I_z^H>0,~I_z^L>0$ and $I_z^H<0,~I_z^L<0$, the FFs' rotation corresponds to the axial (tilting) mode (cooperative rotation), while for $I_z^H>0,~I_z^L<0$ and $I_z^H<0,~I_z^L>0$ it corresponds to the twisting mode (counter rotation)~\cite{Nix1965NP},
as schematically illustrated in the top panel of  Fig.~\ref{fig_AMP_z}.
For the axially symmetric trajectories 1 and 2, the distributions $P(I_z^H,I_z^L)$ are strictly confined to the line  $I_z^H+I_z^L=0$, enforcing exclusively twisting-mode dynamics - i.e., the fragments rotate in opposite directions about the $z$-axis. In contrast, trajectories 3 and 4 (lacking axial symmetry) relax this constraint, allowing both axial and twisting modes to emerge with comparable probabilities.

\begin{figure}[!htbp]
\centering
\includegraphics[width=0.8\textwidth]{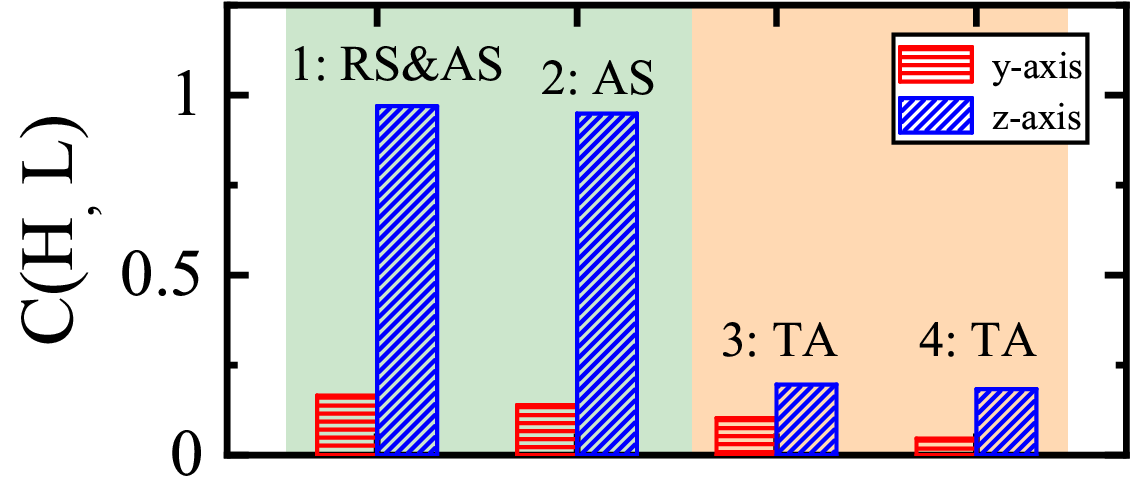}
\caption{(Color online) The correlation indicator $C(H^{i},L^{i})$ between the heavy and light FFs, for projections of their spins
along the $y$- and $z$-axis.}
\label{fig_I}
\end{figure}
%
%\clearpage
To quantify the correlations between the FF spins, we compute the mutual information~$(k=y,z)$: $I(H^{k},L^{k}) = S(H^{k})+S(L^{k}) - S(H^{k},L^{k})$,
where $S(H^{k})$, $S(L^{k})$, and $S(H^{k},L^{k})$ are the information entropies of the projection observables $I_k^H$ and $I_k^L$ (see Supplement~\cite{Li2025Supp} for details).
Mutual information is a measure that quantifies the amount of information obtained about a variable when observing another variable, and it vanishes when there is no correlation between these two variables \cite{Ma2018PPNP}.
Therefore, we define the indicator $C(H^{k},L^{k})=\frac{I(H^{k},L^{k})}{\sqrt{S(H^{k}) S(L^{k})}}$ to quantify the correlations between the spins of FFs.
When $C(H^{k},L^{k})=0$, the spin projections are not correlated, while they are completely correlated when $C(H^{k},L^{k})=1$.
The values of $C(H^{k},L^{k})$ between the projections of FF spins for the trajectories $1-4$ are shown in Fig.~\ref{fig_I}. For projections along the $z$-axis (blue bars),
$C(H^{z},L^{z})$ for trajectories 3 and 4 is only $1/6 - 1/5$ of those computed for the axially symmetric trajectories 1 and 2. For projections along the
$y$-axis (red bars), $C(H^{y},L^{y})$ is considerably smaller in the axial case, while this reduction is less pronounced for the trajectories 3 and 4.

\begin{figure}[!htbp]
\centering
\includegraphics[width=0.8\textwidth]{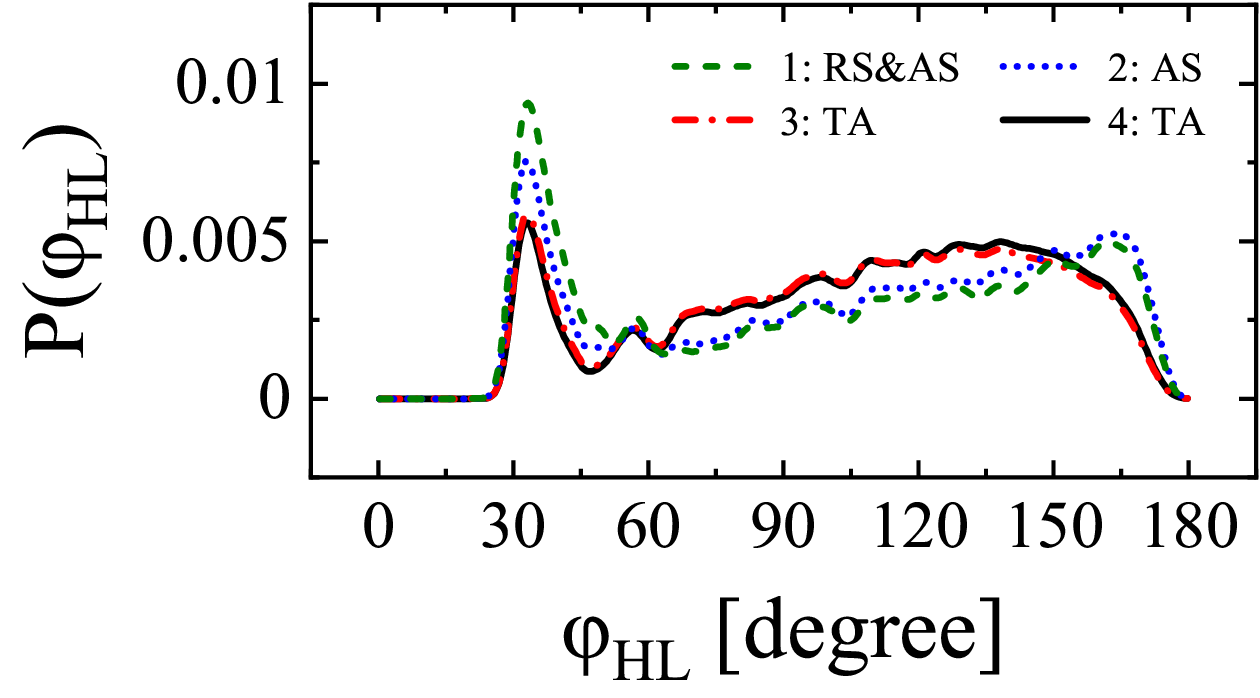}
\caption{(Color online) The probability distributions of the opening angle $\varphi_{HL}$ between the spins of the heavy and light FFs, for the trajectories $1-4$. The computed distributions are smoothed with a Gaussian function of $2^{\circ}$ width.
%The probabilities of the configurations with $\varphi_{HL}=180^{\circ}$ are shown as separate points.
}
\label{fig_OA}
\end{figure}
Next, we consider the opening angle $\varphi_{HL}$ between the spins of the heavy and light FFs. This quantity has been the subject of several recent studies \cite{Bulgac2022PRL,Bulgac2022PRC,Randrup2021PRL,Scamps2022PRC,Scamps2023PRC,Scamps2023PRC2,Scamps2024PRC,Zhou2024}, and the pronounced difference between the predictions of microscopic and phenomenological models for the probability distribution of the opening angle remains an open question. In Fig.~\ref{fig_OA} we plot the probability distributions $P(\varphi_{HL})$ for the trajectories $1-4$. The formulas for computing $P(\varphi_{HL})$ are given in the Supplement~\cite{Li2025Supp}. For the two axially symmetric trajectories (1 and 2), $P(\varphi_{HL})$ essentially reproduces the trend predicted by previous microscopic TDDFT calculations \cite{Scamps2023PRC,Scamps2024PRC}: an almost uniform increase in the interval  $40^\circ \leq \varphi_{HL} \leq 160^\circ$, vanishing $P(\varphi_{HL})$ at small angles $ < 30^\circ$ and around $180^\circ$, with a pronounced peak at $\varphi_{HL} \approx 30^\circ$ and a smaller one at $\varphi_{HL} \approx 165^\circ$. For the two non-axial trajectories (3 and 4), the corresponding distributions $P(\varphi_{HL})$ are essentially identical. Compared to the axial case, the effect of including the triaxial degree of freedom is to considerably reduce the main peak at $\varphi_{HL} \approx 30^\circ$, shift the distribution to larger angles, and smooth out the second peak. The effect is sizeable and demonstrates the importance of including triaxial degrees of freedom in microscopic models of fission dynamics.

Finally, let us briefly discuss the limitations of the framework used here to study spin generation. As noted in the introduction, treating pairing correlations within the BCS approximation neglects the dynamical aspects of the pairing field, which can lead to violations of the continuity equation. More importantly for the spin distribution at scission, the time-dependent mean-field approximation does not account for quantum fluctuations in the collective coordinates. Consequently, each scission event is described by a single fission trajectory. With our recently developed extension of TDDFT - the generalized time-dependent generator coordinate method \cite{Li2023PRC_gdTDGCM,Li2024Fop,Li2025PRC} - a more complete description of fission dynamics becomes possible. This approach explicitly includes fluctuations in the collective degrees of freedom through correlated collective wave functions. Furthermore, symmetries broken at the mean-field level, such as particle number and full angular momentum, need to be restored. However‌, a full dynamical treatment that simultaneously incorporates pairing correlations, quantum fluctuations, and symmetry restoration for non-axial fission geometries remains beyond current computational capabilities.

In conclusion, our microscopic study, based on time-dependent relativistic density functional theory, reveals how triaxial deformation governs spin generation and entanglement in FFs. Breaking axial symmetry broadens the distribution of FF spin projections on the fission axis, enables collective tilting rotations, and reduces spin-spin correlations, as quantified by mutual information. Notably, the opening angle distribution between the spins of the FFs is significantly altered.

\begin{acknowledgments}
This work was supported in part by the High-End Foreign Experts Plan of China,
the National Natural Science Foundation of China (Grants No.12435006, No.12475117, and No.12141501),
the National Key Laboratory of Neutron Science and Technology (Grant No. NST202401016),
the National Key R\&D Program of China 2024YFE0109803,
the National Key Research and Development Program of China 2024YFA1612600,
the High-Performance Computing Platform of Peking University,
by the Postdoctoral Fellowship Program and China Postdoctoral Science Foundation under Grant Number BX20250170, by the project ``Implementation of cutting-edge research and its application as part of the Scientific Center of Excellence for Quantum and Complex Systems, and Representations of Lie Algebras'', PK.1.1.02, European Union, European Regional Development Fund, and by the Croatian Science Foundation under the project Relativistic Nuclear Many-Body Theory in the Multimessenger Observation Era (IP-2022-10-7773).
\end{acknowledgments}

\clearpage
\bigskip
%\begin{CJK*}{GB}{}
\title{Supplemental Material for ``Intrinsic generation of angular momenta and entanglement in fission''}
%\CJKfamily{gbsn}
\author{B. Li}
\affiliation{State Key Laboratory of Nuclear Physics and Technology, School of Physics, Peking University, Beijing 100871, China}
\author{D. D. Zhang}
\affiliation{Institute of Theoretical Physics, Chinese Academy of Science, Beijing 100871, China}
\author{D. Vretenar}
\email{vretenar@phy.hr}
\affiliation{State Key Laboratory of Nuclear Physics and Technology, School of Physics, Peking University, Beijing 100871, China}
\affiliation{Physics Department, Faculty of Science, University of Zagreb, 10000 Zagreb, Croatia}
\author{T. Nik\v si\' c}
\affiliation{Physics Department, Faculty of Science, University of Zagreb, 10000 Zagreb, Croatia}
\author{P. W. Zhao }
\email{pwzhao@pku.edu.cn}
\affiliation{State Key Laboratory of Nuclear Physics and Technology, School of Physics, Peking University, Beijing 100871, China}
\author{J. Meng }
\email{mengj@pku.edu.cn}
\affiliation{State Key Laboratory of Nuclear Physics and Technology, School of Physics, Peking University, Beijing 100871, China}

\date{\today}

\maketitle

This supplemental information contains:
\begin{itemize}
  \item Method: outline of time-dependent relativistic density functional theory used to describe nuclear fission.

  \item Method: definition of deformation parameters.

  \item Figure: deformation energy surface of $^{252}$Cf in the plane of axially symmetric quadrupole $\beta_{20}$ and non-axial octupole $\beta_{32}$ deformation parameters.

  \item Figure: one-body densities of $^{252}$Cf at the initial points of four fission trajectories.

  \item Movies: time evolution of one-body densities along four fission trajectories.

  \item Method: angular momentum projection

  \item Method: mutual information

  \item Method: formula for calculating the opening angle between the intrinsic spins of fission fragments.

  \item Effect of the Coulomb force on angular momenta of fission fragments.
\end{itemize}
%------------------------------------------------------------------------------------------------
\section{Time-dependent density functional theory for nuclear fission}
The time-dependent relativistic density functional theory \cite{Ren2022PRL,Ren2022a} is used to model the time evolution of the quasiparticle vacuum
\begin{equation}
   |\Phi (t)\rangle = \prod_{k>0}[\mu_{k}(t)+\nu_{k}(t)c_{k}^\dagger(t)c_{\bar{k}}^\dagger(t)]|-\rangle \;,
\label{Eq_Slater}
\end{equation}
in the time-dependent BCS approximation~\cite{Ebata2010PRC,Scamps2013PRC}.
In Eq.~(\ref{Eq_Slater}), $\mu_{k}(t)$ and $\nu_{k}(t)$ are the parameters of the transformation between the canonical and quasiparticle bases,
and $c_{k}^\dagger(t)$ denotes the creation operator associated with the canonical state $\phi_k(\bm{r},t)$.~The time evolution of $\phi_k(\bm{r},t)$ is determined by the time-dependent Dirac equation
\begin{equation}\label{Eq_td_Dirac_eq_BCS}
  i\frac{\partial}{\partial t}\phi_k(\bm{r},t)=[\hat{h}(\bm{r},t)-\varepsilon_k(t)]\phi_k(\bm{r},t),
\end{equation}
where $\varepsilon_k(t)=\langle\psi_k|\hat{h}|\psi_k\rangle$ is the single-particle energy, and the single-particle Hamiltonian $\hat{h}(\bm{r},t)$ is determined self-consistently
at each step in time by the time-dependent densities and currents in the scalar, vector, and isovector channels,
\begin{subequations}\label{Eq_density_current}
  \begin{align}
    &\rho_S(\bm{r},t)=\sum_k n_{k}(t)\bar{\phi}_k(\bm{r},t)\phi_k(\bm{r},t),\\
    &j^{\mu}(\bm{r},t)=\sum_k n_{k}(t)\bar{\phi}_k(\bm{r},t)\gamma^\mu\phi_k(\bm{r},t),\\
    &j_{TV}^{\mu}(\bm{r},t)=\sum_k  n_{k}(t)\bar{\phi}_k(\bm{r},t)\gamma^\mu\tau_3\phi_k(\bm{r},t),
  \end{align}
\end{subequations}
where $\tau_3$ is the isospin Pauli matrix, and the vector density reads $\rho_V(\bm{r},t) = j^0(\bm{r},t)$.
The time evolution of the occupation probability $n_{k}(t)=|\nu_{k}(t)|^2$, and pairing tensor $\kappa_{k}(t)=\mu_{k}^{*}(t)\nu_{k}(t)$, is governed by the following equations:
\begin{subequations}\label{Eq_td_nkapp_eq_BCS}
   \begin{align}
     &i\frac{d}{dt}n_{k}(t)=\kappa_{k}(t)\Delta_{k}^*(t)-\kappa_{k}^{*}(t)\Delta_{k}(t),\\
     &i\frac{d}{dt}\kappa_{k}(t)=[\varepsilon_k(t)+\varepsilon_{\bar{k}}(t)]\kappa_{k}(t)+\Delta_{k}(t)[2n_{k}(t)-1] ,
   \end{align}
\end{subequations}
In time-dependent calculations, a monopole pairing interaction is employed, and the gap parameter $\Delta_{k}(t)$ is defined in terms of single-particle energies and the pairing tensor,
\begin{equation}
  \Delta_{k}(t)=\left[G\sum_{k'>0}f(\varepsilon_{k'})\kappa_{k'}\right]f(\varepsilon_k),
\end{equation}
where $f(\varepsilon_k)$ is the cut-off function for the pairing window~\cite{Scamps2013PRC}, and $G$ is the pairing strength.

In the present calculation that employs the static and time-dependent relativistic density functional theory in 3D-coordinate space,
the mesh spacing of the lattice is 1.0 fm for all directions, and the box size is $L_x\times L_y\times L_z=26\times26\times60~{\rm fm}^3$.
The time-dependent Dirac equation \eqref{Eq_td_Dirac_eq_BCS} is solved using the predictor-corrector method, and the time-dependent equations \eqref{Eq_td_nkapp_eq_BCS} with the Euler algorithm.
The step for the time evolution is $0.2~{\rm fm}/c~\approx 6.67\times10^{-4}$~zs.
The pairing strength parameters: -0.125 MeV for neutrons, and -0.210 MeV for protons, are determined by the empirical pairing gaps of $^{252}$Cf, using the three-point odd-even mass formula.
The subspace $V_f$ is selected as $z>0$, where the $z$-axis is along the fission direction, and the interval of $z$ is from $-30$~fm to $30$~fm.
In the computation of the distribution of spin $I$ and its projection $K$ on $z$-axis, the number of Euler angles is $n_\alpha\times n_\beta \times n_\gamma = 20\times 30 \times 20$.
For the joint distribution $P(K_H^k,K_L^k)$, where $K_H^k$ and $K_L^k$ are components of the spin of the heavy and light FFs along the $k$-axis ($k=y,z$),
the number of Euler angles is $n_{\theta_H}\times n_{\theta_L} = 200\times 200$.

\clearpage

\section{Deformation parameters}
For a nucleon (vector) density distribution $\rho_V(\bm r)$ determined in a self-consistent relativistic mean-field calculation, the intrinsic multipole deformation parameters are defined:
\begin{equation}
\beta_{\lambda\mu} = \frac{2\pi}{3AR_0^\lambda} \int d {\bm r} \rho_V(\bm r)r^\lambda
\left\{
\begin{aligned}
 &[Y_{\lambda\mu}(\Omega)+(-1)^{\mu}Y_{\lambda-\mu}(\Omega)],~~~\mu\geq 0 \\
&i[Y_{\lambda-\mu}(\Omega)(-1)^{\mu} -Y_{\lambda\mu}(\Omega)],~~~\mu< 0
\end{aligned}
\right.,
\end{equation}
where $Y_{\lambda\mu}(\Omega)$ denotes spherical harmonics, $A$ is the mass number of the nucleus, and $R_0 =1.2 A^{1/3}~{\rm fm}$.
Note that spherical harmonics satisfy the equation $Y_{\lambda\mu}^{*}(\Omega) = (-1)^\mu Y_{\lambda-\mu}(\Omega)$, and all deformation parameters $\beta_{\lambda\mu}$ are real numbers.

\section{Deformation energy surface of $^{252}$Cf in the ($\beta_{20}$-$\beta_{32}$) plane}
\begin{figure}[!htbp]
\centering
\includegraphics[width=0.8\textwidth]{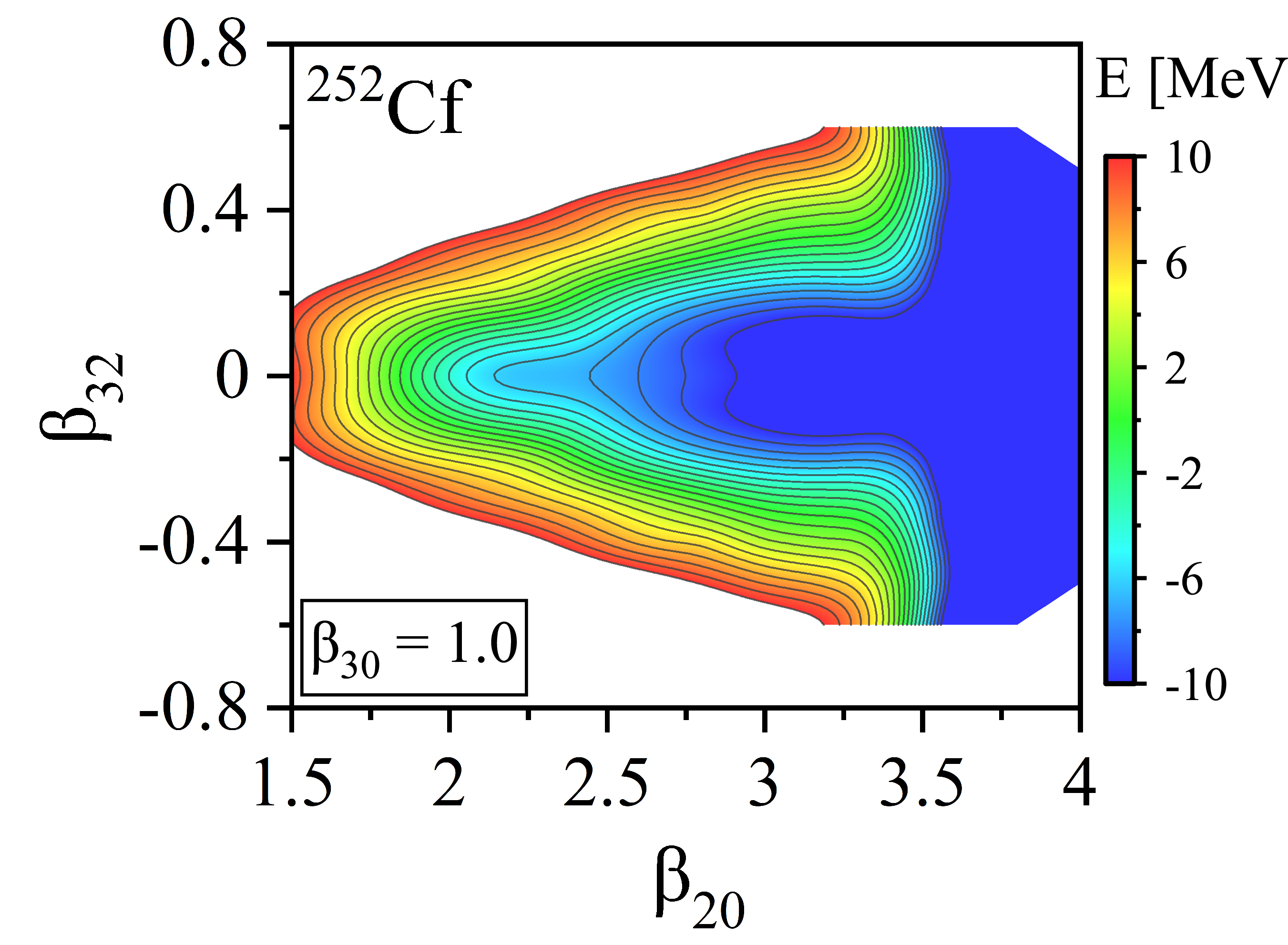}
\caption{(Color online) Self-consistent deformation energy surface of $^{252}$Cf in the plane of axially symmetric quadrupole $\beta_{20}$ and non-axial octupole $\beta_{32}$ deformation parameters, for a fixed value $\beta_{30} = 1.0$,
computed with the relativistic density functional PC-PK1~\cite{Zhao2010PRC} and a monopole pairing interaction. Contours join points on the surface with the same energy.}
\label{fig_ES}
\end{figure}

\clearpage
\section{One-body densities of $^{252}$Cf at the initial points of four fission trajectories}
\begin{figure}[!htbp]
\centering
\includegraphics[width=0.8\textwidth]{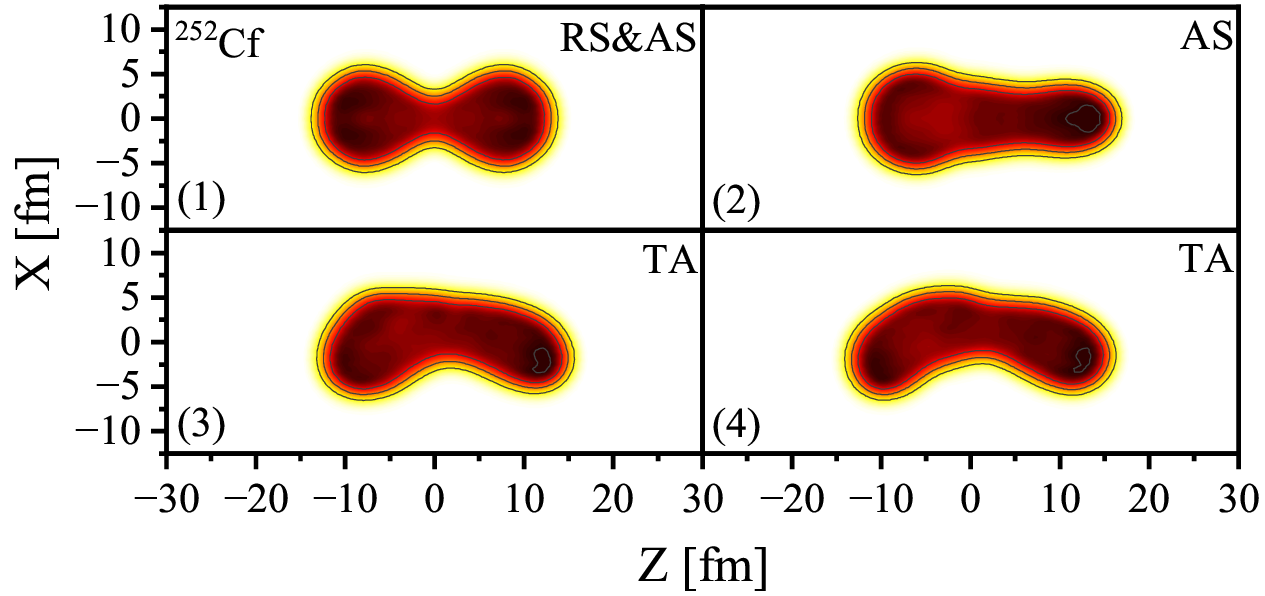}
\caption{(Color online) Panels (1)-(4): The one-body densities of $^{252}$Cf at the initial points of the fission trajectories $1-4$,
shown in Fig. 1 of the article.}
\label{fig_initial}
\end{figure}
\section{Time evolution of one-body densities along four fission trajectories}
The movies 2Drho\_tr1-4.gif display the time evolution of one-body densities in the $x$-$z$ coordinate plane for trajectories 1-4, shown in Fig. 1 of the article.

\clearpage
\section{Angular momentum projection}
For each fragment, the probability distributions $P(I,I_z)$ of the spin $I$ and its projection $I_z$ on the $z$-axis, are
obtained by angular momentum projection:
\begin{equation}
\begin{aligned}
P(I,I_z) &= \langle\Psi|\hat{P}^I_{I_zI_z}|\Psi\rangle\\
&=\frac{2I+1}{16\pi^2} \int_0^{2\pi}d\alpha \int_0^{\pi} \sin\beta d\beta \int_0^{4\pi}d\gamma  \\
&D^{I*}_{I_zI_z}(\alpha,\beta,\gamma)\langle\Psi|e^{-(i\hat{J}_z\alpha+i\hat{J}_y\beta+i\hat{J}_z\gamma)\Theta_{V_f}({\bm r})}|\Psi\rangle,
\end{aligned}
\label{PIK}
\end{equation}
where $D^{I}_{I_zI_z}(\alpha,\beta,\gamma)$ is the Wigner-D function. The Heaviside function $\Theta_{V_f}({\bm r})$ divides the space into the volume $V_f~(f=H, {\rm or}~L)$ in which the fragment is located, and the complementary volume.

To evaluate correlations between the intrinsic angular momenta of the two FFs, we compute the joint distribution
$P(I_k^H,I_k^L)$, where $I_k^H$ and $I_k^L$ are components of the spin of the heavy and light FFs along the $k$-axis, respectively:
\begin{equation}
\begin{aligned}
    P(I_k^H,I_k^L) &= \langle\Psi|\hat{P}_{I_k^H}\hat{P}_{I_k^L}|\Psi\rangle\\
    &=\frac{1}{16\pi^2} \int_0^{4\pi}d\theta_H\int_0^{4\pi}d\theta_L e^{i\theta_H I_k^H+i\theta_L I_k^L}\\
    &\langle\Psi|e^{-i\theta_H \hat{J}_k\Theta_{V_H}({\bm r})}e^{-i\theta_L \hat{J}_k\Theta_{V_L}({\bm r})}|\Psi\rangle.
\end{aligned}
\label{PKK}
\end{equation}

\clearpage
\section{Mutual information}
To quantify the correlations between the FF spins, we compute the mutual information~$(k=y,z)$,
\begin{equation}
    I(H^{k},L^{k}) = S(H^{k})+S(L^{k}) - S(H^{k},L^{k}).
\end{equation}
where $S(H^{k})$, $S(L^{k})$, and $S(H^{k},L^{k})$ are the information entropies of the projection observables $I_k^H$ and $I_k^L$:
\begin{equation}
\begin{aligned}
 &S(f^{k}) = -\sum_{I^f_k} P(I^f_k) \ln P(I^f_k), ~~~f = H, L\\
 &S(H^{k},L^{k}) = -\sum_{I_k^H,I_k^L} P(I_k^H,I_k^L) \ln  P(I_k^H,I_k^L).
\end{aligned}
\end{equation}
$P(I^f_k) = \sum_{I^{f'}_k} P(I^f_k,I^{f'}_k)$ [$(f,f')=(H,L)$ or $(L,H)$] and $S(H^{k},L^{k})$ are computed from the distributions of Eq.~(\ref{PKK}).

\clearpage
\section{Formula for the opening angle between FF intrinsic spins}
The probability distribution $P(\varphi_{HL})$ of the opening angle is evaluated from the triple distribution $P(\Lambda,I_H,I_L)$~\cite{Scamps2024PRC},
\begin{equation}
P(\varphi_{HL}) = \sum_{\Lambda,I_H\neq0,I_L\neq0} \Theta(|\varphi_{HL}-\varphi_{HL}^{'}(\Lambda,I_H,I_L)|)P(\Lambda,I_H,I_L),
\end{equation}
where the Heaviside step function $\Theta(x)$ is defined
\begin{equation}
\Theta(x)=\left\{
\begin{aligned}
&1, x \leq \sigma \\
&0, x > \sigma
\end{aligned}
\right.,
\end{equation}
and for  $\sigma$ the value  $0.5^{\circ}$ is used in the calculation considered in the present study.
The opening angle $\varphi_{HL}^{'}(\Lambda,I_H,I_L)$ reads
\begin{equation}
\varphi_{HL}^{'}(\Lambda,I_H,I_L) = \arccos \left ( \frac{\Lambda(\Lambda+1)-I_H(I_H+1)-I_L(I_L+1)}{2\sqrt{I_H(I_H+1)I_L(I_L+1)}}\right ).
\end{equation}
The triple distribution $P(\Lambda,I_H,I_L)$ is computed from
\begin{equation}
P(\Lambda,I_H,I_L) = \sum_{\Lambda_z,I^H_z,I^L_z}\langle\Psi|\hat{P}^{\Lambda}_{\Lambda_z\Lambda_z}\hat{P}^{I_H}_{I^H_z I^H_z}, \hat{P}^{I_L}_{I^L_z I^L_z}|\Psi\rangle
\label{eq_P_tri}
\end{equation}
where the angular momentum projections are defined as
\begin{equation}
\hat{P}^{\Lambda}_{\Lambda_z\Lambda^{'}_z} = \frac{2\Lambda+1}{8\pi^2} \int d\Omega~D^{\Lambda,*}_{\Lambda_z\Lambda^{'}_z}(\Omega) \hat{R}(\Omega).
\end{equation}
\begin{equation}
\hat{P}^{I_f}_{I^f_z M^f} = \frac{2I_f+1}{16\pi^2} \int d\Omega_f~D^{I_f,*}_{I^f_z M^f}(\Omega_f) \hat{R}^f(\Omega_f),~~~~f = H,L.
\end{equation}
The operator $\hat{R}(\Omega)$ rotates the fission direction, and is performed in the reference frame of the center of mass of the total system.
The rotation center of operator $\hat{R}^f(\Omega_f)$ is the center of mass of the FF in the subspace $V_f$.
The formula for calculating the triple distribution $P(\Lambda,I_H,I_L)$ is same as that introduced in the supplement of Ref.~\cite{Scamps2023PRC} but,
for non-axial trajectories, without the constraint $\Lambda_z = I^H_z +I^L_z = 0$.
Eq.~(\ref{eq_P_tri}) reduces then to the following projection formula for the triple distribution
\begin{equation}
\begin{aligned}
&P(\Lambda,I_H,I_L) =\\
&\sum_{\Lambda_z,I^H_z,M_H,I^L_z,M_L} (-1)^{I^H_z-M_H+I^L_z-M_L} C^{\Lambda,\Lambda_z}_{I_H,-I^H_z,I_L,-I^L_z}C^{\Lambda,\Lambda_z}_{I_H,-M_H,I_L,-M_L}\langle\Psi|\hat{P}^{I_H}_{I^H_zM_H}\hat{P}^{I_L}_{I^L_zM_L}|\Psi\rangle,
\end{aligned}
\end{equation}
where $C^{JM}_{j_1,m_1,j_2,m_2}$ is the Clebsch-Gordan coefficient.

The matrix element $\langle\Psi|\hat{P}^{I_H}_{I^H_zM_H}\hat{P}^{I_L}_{I^L_zM_L}|\Psi\rangle$ reads
\begin{equation}
\langle\Psi|\hat{P}^{I_H}_{I^H_zM_H}\hat{P}^{I_L}_{I^L_zM_L}|\Psi\rangle = \int d\Omega_H \int d\Omega_L~
D^{I_H,*}_{I^H_z M^H}(\Omega_H)D^{I_L,*}_{I^L_z M^L}(\Omega_L)\langle\Psi|\hat{R}^H(\Omega_H)\hat{R}^L(\Omega_L)|\Psi\rangle,
\end{equation}
and the rotational matrix element can be calculated using the Pfaffian method~\cite{Hu2014PLB},
\begin{equation}
\langle \Psi|\hat{R}^H(\Omega_H)\hat{R}^L(\Omega_L)|\Psi\rangle = \langle -|\prod_{i>0}(\mu_i^{*}+\nu_i^{*}a_{\bar{i}}a_i )\prod_{j>0}(\mu_j+\nu_jb_j^\dagger b_{\bar{j}}^\dagger)|-\rangle=
{\rm Pf}[S(\Omega_H,\Omega_L)].
\end{equation}
where the operators $a_i$ and $b_j^\dagger$ satisfy
\begin{equation}
    a_i = \hat{R}^{H,\dagger}(\Omega_H) c_i \hat{R}^H(\Omega_H),
\end{equation}
\begin{equation}
    b_j^\dagger = \hat{R}^{L}(\Omega_L) c_j^\dagger \hat{R}^{L,\dagger}(\Omega_L).
\end{equation}
$c^{\dagger}$ is the creation operator for the canonical states defined in Eq.~(\ref{Eq_Slater}).\\
The matrix $S(\Omega_H,\Omega_L)$ is defined
\begin{equation}
    S(\Omega_H,\Omega_L)=
    \left[\begin{array}{cc}
    S_a & S_{ab}\\
    -S_{ab}^T & S_b \\
    \end{array}\right] ,
\end{equation}
with the elements
\begin{equation}
    (S_a)_{ij}=-\nu_i^{*}\mu_j^{*}\langle-|a_{\bar{i}}a^{\dagger}_j|-\rangle =-\nu_i^{*}\mu_j^{*}\delta_{\bar{i}j},
\end{equation}
\begin{equation}
    (S_b)_{ij}=-\mu_i\nu_j\langle-|b_ib^{\dagger}_{\bar{j}}|-\rangle=-\mu_i\nu_j\delta_{i\bar{j}},
\end{equation}
\begin{equation}
    (S_{ab})_{ij}=\nu_i^{*}\nu_j\langle-|a_{\bar{i}}b_{\bar{j}}^\dagger|-\rangle=\nu_i^{*}\nu_j [\langle\phi_{\bar{i}}|\hat{R}^H(\Omega_H)|\phi_{\bar{j}}\rangle_{V_H} + \langle\phi_{\bar{i}}|\hat{R}^L(\Omega_L)|\phi_{\bar{j}}\rangle_{V_L}].
\end{equation}
The rotation matrix elements between canonical states in the subspace $V_f~(f = H$ or $L)$ read
\begin{equation}
\langle\phi_i|\hat{R}^f(\Omega_f)|\phi_j\rangle_{V_f} = \int d {\bm r}~\Theta_{V_f}({\bm r}) \phi_i^{\dagger}(\bm r) \hat{R}^f(\Omega_f) \phi_j(\bm r),
\end{equation}
where the Heaviside function $\Theta_{V_f}({\bm r})$ divides the space into the volume $V_f$ in which the fragment is located, and the complementary volume.

The matrix element $\langle\phi_i|\hat{R}^f(\Omega_f)|\phi_j\rangle_{V_f} $ can be rewritten as
\begin{equation}
    \langle\phi_i|\hat{R}^f(\Omega_f)|\phi_j\rangle_{V_f} = \langle\phi_i|e^{-i\hat{J}_z\alpha_f}e^{-i\hat{J}_y\beta_f}e^{-i\hat{J}_z\gamma_f}|\phi_j\rangle_{V_f}
    =\langle \tilde{\phi}_i(\alpha_f)|\tilde{\phi}_j(\beta_f,\gamma_f)\rangle_{V_f},
\end{equation}
where $|\tilde{\phi}_i(\alpha_f)\rangle$ and $|\tilde{\phi}_j(\beta_f,\gamma_f)\rangle$ read
\begin{equation}
    |\tilde{\phi}_i(\alpha_f)\rangle = e^{i\hat{J}_z\alpha_f} |\phi_i\rangle,
\end{equation}
\begin{equation}
    |\tilde{\phi}_j(\beta_f,\gamma_f)\rangle = e^{-i\hat{J}_y\beta_f}e^{-i\hat{J}_z\gamma_f}|\phi_j\rangle.
\end{equation}
$e^{i\hat{J}_z\alpha_f}$, $e^{-i\hat{J}_y\beta_f}$, and $e^{-i\hat{J}_z\gamma_f}$ are the rotation operators for the FF in the subspace $V_f$.
The overlap between the coordinate basis $|{\bm r}\rangle$ and $ |\tilde{\phi}_i(\alpha_f)\rangle$ in subspace $V_f$ can be obtained by
\begin{equation}
     \langle {\bm r}|\tilde{\phi}_i(\alpha_f)\rangle_{V_f} =  \langle {\bm r}|e^{i\hat{J}_z\alpha_f}|\phi_i\rangle_{V_f} = \langle {\bm r'}(\alpha_f)|\phi_i\rangle_{V_f},
\end{equation}
where $|{\bm r'}(\alpha_f)\rangle = e^{-i\hat{J}_z\alpha_f}|{\bm r}\rangle$, and their coordinate components satisfy
\begin{equation}
    x' = x\cos(\alpha_f)-y\sin(\alpha_f);~~y' = y\cos(\alpha_f)+x\sin(\alpha_f);~~z'=z.
\end{equation}
Note that $\langle {\bm r}|\tilde{\phi}_i(\alpha_f)\rangle_{V_f}$ are stored for repeated use in numerical calculations.
We can calculate $\langle {\bm r}|\tilde{\phi}_j(\beta_f,\gamma_f)\rangle_{V_f}$ with the same method,
and compute the overlaps between $|\tilde{\phi}_j(\beta_f,\gamma_f)\rangle $ and $|\tilde{\phi}_i(\alpha_f)\rangle$ in the subspace $V_f$.

The matrix elements $\langle\phi_i|\hat{R}^f(\Omega_f)|\phi_j\rangle_{V_f}, (f = H,L)$ are calculated independently and stored.
When computing the Pfaffian of the matrix $S(\Omega_H,\Omega_L)$, the corresponding $\langle\phi_i|\hat{R}^f(\Omega_f)|\phi_j\rangle_{V_f}$ are directly read from the storage medium.
In addition, we note that the calculations for different Euler angles can be carried out in parallel.

\section{The effect of Coulomb force on angular momenta of fission fragments}
\begin{figure}[!htbp]
\centering
\includegraphics[width=0.8\textwidth]{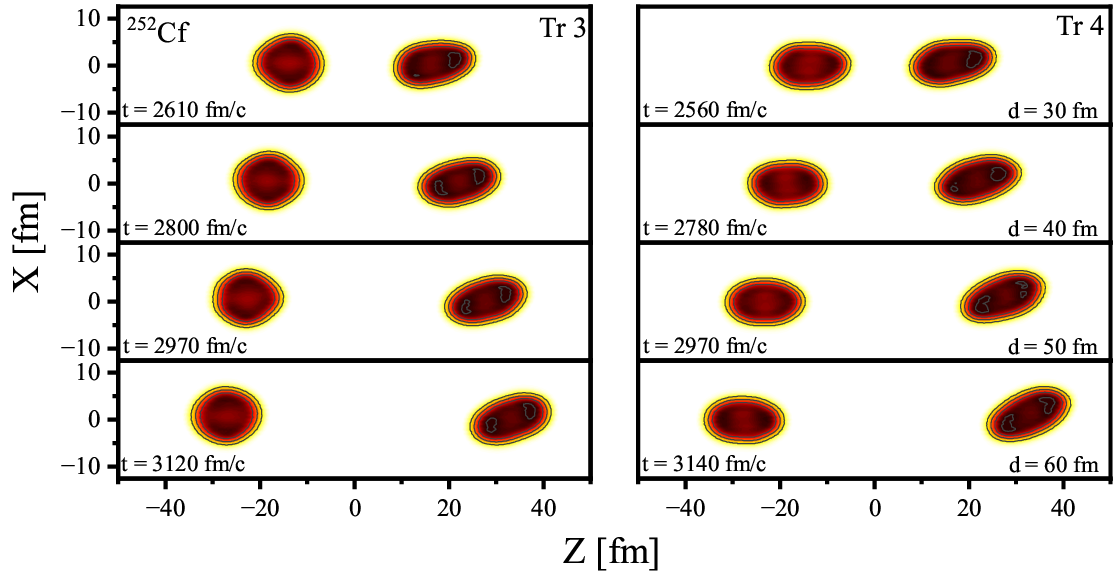}
\caption{Left panel: The one-body densities of FFs along trajectory 3 after spontaneous fission of $^{252}$Cf, when the distance $d$ between the centers of mass of two fragments equals: $30$, $40$, $50$, and $60$ fm.
Right panel: Same as in caption to the left panel but for trajectory 4. }
\label{fig_dense_t}
\end{figure}

\begin{figure}[!htbp]
\centering
\includegraphics[width=0.8\textwidth]{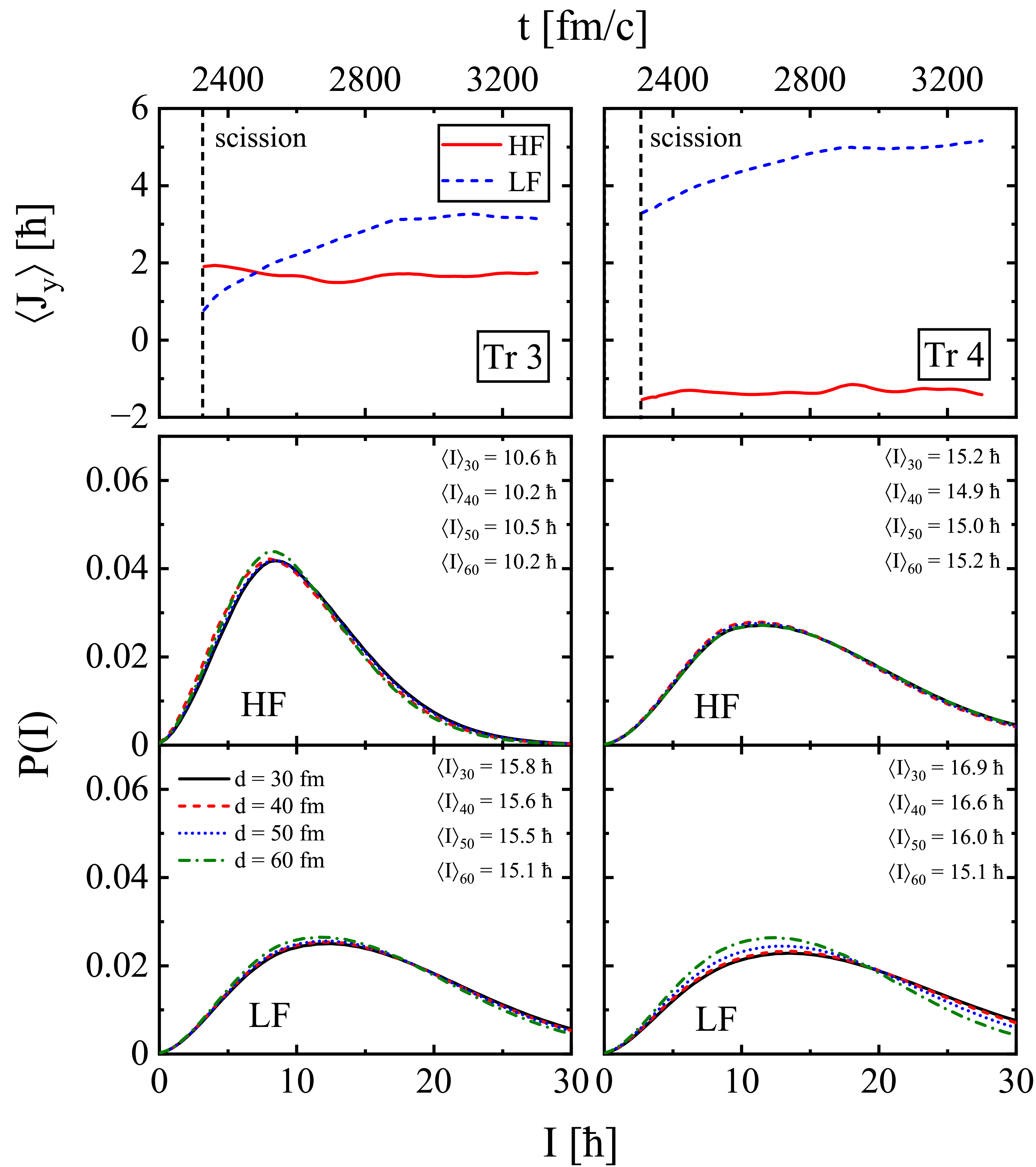}
\caption{The first row shows the time evolution of the expectation value of the FF angular momentum projection along the $y$-axis for trajectories 3 (left panel) and 4 (right panel). In the second and third rows we display the corresponding probability distributions of the angular momentum $I$ of the heavy and light fragments, respectively,
when the distance $d$ between the centers of mass of the two fragments is: $30$, $40$, $50$, and $60$ fm.
The insets show the average angular momenta at these distances, obtained from the corresponding probability distributions.
}
\label{fig_Coulomb}
\end{figure}

The expectation value of the angular momentum projection along the $i$-axis reads~\cite{Scamps2022PRC},
\begin{equation}
\langle J_i\rangle = {\bm l_i\cdot} \sum_k n_k \langle \phi_k| [(\hat{{\bm r}}-{\bm r_{cm}})\times ({\hat{{\bm p}}}-{\bm p_{cm}})+\hat{{\bm s}}]\Theta_{V_f}({\bm r})|\phi_k\rangle.
\end{equation}
${\bm r_{cm}}$ and ${\bm p_{cm}}$ are the position and momentum of the center of mass of the fragment, respectively.
The Heaviside function $\Theta_{V_f}({\bm r})$ divides the space into the volume $V_f~(f = H$ or $L)$ in which the fragment is located, and the complementary volume.
$l_i$ is the unit vector corresponding to the $i$-axis. The numerical details for the calculations illustrated in Figs.~\ref{fig_dense_t} and \ref{fig_Coulomb}
are the same as Fig. 2 in the manuscript, except that the box size is extended in the $z$-direction ($L_x\times L_y\times L_z=26\times26\times100~{\rm fm}^3$).

\clearpage
%\bibliography{paper}

\clearpage

%\bibliography{supplement}

\begin{thebibliography}{73}%
\makeatletter
\providecommand \@ifxundefined [1]{%
 \@ifx{#1\undefined}
}%
\providecommand \@ifnum [1]{%
 \ifnum #1\expandafter \@firstoftwo
 \else \expandafter \@secondoftwo
 \fi
}%
\providecommand \@ifx [1]{%
 \ifx #1\expandafter \@firstoftwo
 \else \expandafter \@secondoftwo
 \fi
}%
\providecommand \natexlab [1]{#1}%
\providecommand \enquote  [1]{``#1''}%
\providecommand \bibnamefont  [1]{#1}%
\providecommand \bibfnamefont [1]{#1}%
\providecommand \citenamefont [1]{#1}%
\providecommand \href@noop [0]{\@secondoftwo}%
\providecommand \href [0]{\begingroup \@sanitize@url \@href}%
\providecommand \@href[1]{\@@startlink{#1}\@@href}%
\providecommand \@@href[1]{\endgroup#1\@@endlink}%
\providecommand \@sanitize@url [0]{\catcode `\\12\catcode `\$12\catcode
  `\&12\catcode `\#12\catcode `\^12\catcode `\_12\catcode `\%12\relax}%
\providecommand \@@startlink[1]{}%
\providecommand \@@endlink[0]{}%
\providecommand \url  [0]{\begingroup\@sanitize@url \@url }%
\providecommand \@url [1]{\endgroup\@href {#1}{\urlprefix }}%
\providecommand \urlprefix  [0]{URL }%
\providecommand \Eprint [0]{\href }%
\providecommand \doibase [0]{http://dx.doi.org/}%
\providecommand \selectlanguage [0]{\@gobble}%
\providecommand \bibinfo  [0]{\@secondoftwo}%
\providecommand \bibfield  [0]{\@secondoftwo}%
\providecommand \translation [1]{[#1]}%
\providecommand \BibitemOpen [0]{}%
\providecommand \bibitemStop [0]{}%
\providecommand \bibitemNoStop [0]{.\EOS\space}%
\providecommand \EOS [0]{\spacefactor3000\relax}%
\providecommand \BibitemShut  [1]{\csname bibitem#1\endcsname}%
\let\auto@bib@innerbib\@empty
%</preamble>
\bibitem [{\citenamefont {Schmidt}\ and\ \citenamefont
  {Jurado}(2018)}]{Schmidt2018RPP}%
  \BibitemOpen
  \bibfield  {author} {\bibinfo {author} {\bibfnamefont {K.-H.}\ \bibnamefont
  {Schmidt}}\ and\ \bibinfo {author} {\bibfnamefont {B.}~\bibnamefont
  {Jurado}},\ }\href {\doibase 10.1088/1361-6633/aacfa7} {\bibfield  {journal}
  {\bibinfo  {journal} {Rep. Prog. Phys.}\ }\textbf {\bibinfo {volume} {81}},\
  \bibinfo {pages} {106301} (\bibinfo {year} {2018})}\BibitemShut {NoStop}%
\bibitem [{\citenamefont {Bender}\ \emph {et~al.}(2020)\citenamefont {Bender},
  \citenamefont {Bernard}, \citenamefont {Bertsch}, \citenamefont {Chiba},
  \citenamefont {Dobaczewski}, \citenamefont {Dubray}, \citenamefont
  {Giuliani}, \citenamefont {Hagino}, \citenamefont {Lacroix}, \citenamefont
  {Li}, \citenamefont {Magierski}, \citenamefont {Maruhn}, \citenamefont
  {Nazarewicz}, \citenamefont {Pei}, \citenamefont {P{\'{e}}ru}, \citenamefont
  {Pillet}, \citenamefont {Randrup}, \citenamefont {Regnier}, \citenamefont
  {Reinhard}, \citenamefont {Robledo}, \citenamefont {Ryssens}, \citenamefont
  {Sadhukhan}, \citenamefont {Scamps}, \citenamefont {Schunck}, \citenamefont
  {Simenel}, \citenamefont {Skalski}, \citenamefont {Stetcu}, \citenamefont
  {Stevenson}, \citenamefont {Umar}, \citenamefont {Verriere}, \citenamefont
  {Vretenar}, \citenamefont {Warda},\ and\ \citenamefont
  {{\AA}berg}}]{Bender2020}%
  \BibitemOpen
  \bibfield  {author} {\bibinfo {author} {\bibfnamefont {M.}~\bibnamefont
  {Bender}}, \bibinfo {author} {\bibfnamefont {R.}~\bibnamefont {Bernard}},
  \bibinfo {author} {\bibfnamefont {G.}~\bibnamefont {Bertsch}}, \bibinfo
  {author} {\bibfnamefont {S.}~\bibnamefont {Chiba}}, \bibinfo {author}
  {\bibfnamefont {J.}~\bibnamefont {Dobaczewski}}, \bibinfo {author}
  {\bibfnamefont {N.}~\bibnamefont {Dubray}}, \bibinfo {author} {\bibfnamefont
  {S.~A.}\ \bibnamefont {Giuliani}}, \bibinfo {author} {\bibfnamefont
  {K.}~\bibnamefont {Hagino}}, \bibinfo {author} {\bibfnamefont
  {D.}~\bibnamefont {Lacroix}}, \bibinfo {author} {\bibfnamefont
  {Z.}~\bibnamefont {Li}}, \bibinfo {author} {\bibfnamefont {P.}~\bibnamefont
  {Magierski}}, \bibinfo {author} {\bibfnamefont {J.}~\bibnamefont {Maruhn}},
  \bibinfo {author} {\bibfnamefont {W.}~\bibnamefont {Nazarewicz}}, \bibinfo
  {author} {\bibfnamefont {J.}~\bibnamefont {Pei}}, \bibinfo {author}
  {\bibfnamefont {S.}~\bibnamefont {P{\'{e}}ru}}, \bibinfo {author}
  {\bibfnamefont {N.}~\bibnamefont {Pillet}}, \bibinfo {author} {\bibfnamefont
  {J.}~\bibnamefont {Randrup}}, \bibinfo {author} {\bibfnamefont
  {D.}~\bibnamefont {Regnier}}, \bibinfo {author} {\bibfnamefont {P.-G.}\
  \bibnamefont {Reinhard}}, \bibinfo {author} {\bibfnamefont {L.~M.}\
  \bibnamefont {Robledo}}, \bibinfo {author} {\bibfnamefont {W.}~\bibnamefont
  {Ryssens}}, \bibinfo {author} {\bibfnamefont {J.}~\bibnamefont {Sadhukhan}},
  \bibinfo {author} {\bibfnamefont {G.}~\bibnamefont {Scamps}}, \bibinfo
  {author} {\bibfnamefont {N.}~\bibnamefont {Schunck}}, \bibinfo {author}
  {\bibfnamefont {C.}~\bibnamefont {Simenel}}, \bibinfo {author} {\bibfnamefont
  {J.}~\bibnamefont {Skalski}}, \bibinfo {author} {\bibfnamefont
  {I.}~\bibnamefont {Stetcu}}, \bibinfo {author} {\bibfnamefont
  {P.}~\bibnamefont {Stevenson}}, \bibinfo {author} {\bibfnamefont
  {S.}~\bibnamefont {Umar}}, \bibinfo {author} {\bibfnamefont {M.}~\bibnamefont
  {Verriere}}, \bibinfo {author} {\bibfnamefont {D.}~\bibnamefont {Vretenar}},
  \bibinfo {author} {\bibfnamefont {M.}~\bibnamefont {Warda}}, \ and\ \bibinfo
  {author} {\bibfnamefont {S.}~\bibnamefont {{\AA}berg}},\ }\href {\doibase
  10.1088/1361-6471/abab4f} {\bibfield  {journal} {\bibinfo  {journal} {J.
  Phys. G: Nucl. Part. Phys.}\ }\textbf {\bibinfo {volume} {47}},\ \bibinfo
  {pages} {113002} (\bibinfo {year} {2020})}\BibitemShut {NoStop}%
\bibitem [{\citenamefont {Schunck}\ and\ \citenamefont
  {Regnier}(2022)}]{Schunck22PPNP}%
  \BibitemOpen
  \bibfield  {author} {\bibinfo {author} {\bibfnamefont {N.}~\bibnamefont
  {Schunck}}\ and\ \bibinfo {author} {\bibfnamefont {D.}~\bibnamefont
  {Regnier}},\ }\href {\doibase https://doi.org/10.1016/j.ppnp.2022.103963}
  {\bibfield  {journal} {\bibinfo  {journal} {Prog. Part. Nucl. Phys.}\
  }\textbf {\bibinfo {volume} {125}},\ \bibinfo {pages} {103963} (\bibinfo
  {year} {2022})}\BibitemShut {NoStop}%
\bibitem [{\citenamefont {Nix}\ and\ \citenamefont
  {Swiatecki}(1965)}]{Nix1965NP}%
  \BibitemOpen
  \bibfield  {author} {\bibinfo {author} {\bibfnamefont {J.~R.}\ \bibnamefont
  {Nix}}\ and\ \bibinfo {author} {\bibfnamefont {W.~J.}\ \bibnamefont
  {Swiatecki}},\ }\href {\doibase https://doi.org/10.1016/0029-5582(65)90038-6}
  {\bibfield  {journal} {\bibinfo  {journal} {Nucl. Phys.}\ }\textbf {\bibinfo
  {volume} {71}},\ \bibinfo {pages} {1} (\bibinfo {year} {1965})}\BibitemShut
  {NoStop}%
\bibitem [{\citenamefont {Rasmussen}\ \emph {et~al.}(1969)\citenamefont
  {Rasmussen}, \citenamefont {Nörenberg},\ and\ \citenamefont
  {Mang}}]{Rasmussen1969NPA}%
  \BibitemOpen
  \bibfield  {author} {\bibinfo {author} {\bibfnamefont {J.}~\bibnamefont
  {Rasmussen}}, \bibinfo {author} {\bibfnamefont {W.}~\bibnamefont
  {Nörenberg}}, \ and\ \bibinfo {author} {\bibfnamefont {H.}~\bibnamefont
  {Mang}},\ }\href {\doibase https://doi.org/10.1016/0375-9474(69)90066-9}
  {\bibfield  {journal} {\bibinfo  {journal} {Nucl. Phys. A}\ }\textbf
  {\bibinfo {volume} {136}},\ \bibinfo {pages} {465} (\bibinfo {year}
  {1969})}\BibitemShut {NoStop}%
\bibitem [{\citenamefont {Zielinska-Pfabé}\ and\ \citenamefont
  {Dietrich}(1974)}]{ZielinskaPfabe1974PLB}%
  \BibitemOpen
  \bibfield  {author} {\bibinfo {author} {\bibfnamefont {M.}~\bibnamefont
  {Zielinska-Pfabé}}\ and\ \bibinfo {author} {\bibfnamefont {K.}~\bibnamefont
  {Dietrich}},\ }\href {\doibase https://doi.org/10.1016/0370-2693(74)90488-2}
  {\bibfield  {journal} {\bibinfo  {journal} {Phys. Lett. B}\ }\textbf
  {\bibinfo {volume} {49}},\ \bibinfo {pages} {123} (\bibinfo {year}
  {1974})}\BibitemShut {NoStop}%
\bibitem [{\citenamefont {Bonneau}\ \emph {et~al.}(2007)\citenamefont
  {Bonneau}, \citenamefont {Quentin},\ and\ \citenamefont
  {Mikhailov}}]{Bonneau2007PRC}%
  \BibitemOpen
  \bibfield  {author} {\bibinfo {author} {\bibfnamefont {L.}~\bibnamefont
  {Bonneau}}, \bibinfo {author} {\bibfnamefont {P.}~\bibnamefont {Quentin}}, \
  and\ \bibinfo {author} {\bibfnamefont {I.~N.}\ \bibnamefont {Mikhailov}},\
  }\href {\doibase 10.1103/PhysRevC.75.064313} {\bibfield  {journal} {\bibinfo
  {journal} {Phys. Rev. C}\ }\textbf {\bibinfo {volume} {75}},\ \bibinfo
  {pages} {064313} (\bibinfo {year} {2007})}\BibitemShut {NoStop}%
\bibitem [{\citenamefont {Wilhelmy}\ \emph {et~al.}(1972)\citenamefont
  {Wilhelmy}, \citenamefont {Cheifetz}, \citenamefont {Jared}, \citenamefont
  {Thompson}, \citenamefont {Bowman},\ and\ \citenamefont
  {Rasmussen}}]{Wilhelmy1972PRC}%
  \BibitemOpen
  \bibfield  {author} {\bibinfo {author} {\bibfnamefont {J.~B.}\ \bibnamefont
  {Wilhelmy}}, \bibinfo {author} {\bibfnamefont {E.}~\bibnamefont {Cheifetz}},
  \bibinfo {author} {\bibfnamefont {R.~C.}\ \bibnamefont {Jared}}, \bibinfo
  {author} {\bibfnamefont {S.~G.}\ \bibnamefont {Thompson}}, \bibinfo {author}
  {\bibfnamefont {H.~R.}\ \bibnamefont {Bowman}}, \ and\ \bibinfo {author}
  {\bibfnamefont {J.~O.}\ \bibnamefont {Rasmussen}},\ }\href {\doibase
  10.1103/PhysRevC.5.2041} {\bibfield  {journal} {\bibinfo  {journal} {Phys.
  Rev. C}\ }\textbf {\bibinfo {volume} {5}},\ \bibinfo {pages} {2041} (\bibinfo
  {year} {1972})}\BibitemShut {NoStop}%
\bibitem [{\citenamefont {Wolf}\ and\ \citenamefont
  {Cheifetz}(1976)}]{Wolf1976PRC}%
  \BibitemOpen
  \bibfield  {author} {\bibinfo {author} {\bibfnamefont {A.}~\bibnamefont
  {Wolf}}\ and\ \bibinfo {author} {\bibfnamefont {E.}~\bibnamefont
  {Cheifetz}},\ }\href {\doibase 10.1103/PhysRevC.13.1952} {\bibfield
  {journal} {\bibinfo  {journal} {Phys. Rev. C}\ }\textbf {\bibinfo {volume}
  {13}},\ \bibinfo {pages} {1952} (\bibinfo {year} {1976})}\BibitemShut
  {NoStop}%
\bibitem [{\citenamefont {Wilson}\ \emph {et~al.}(2021)\citenamefont {Wilson},
  \citenamefont {Thisse}, \citenamefont {Lebois}, \citenamefont {Jovan\ifmmode
  \check{c}\else \v{c}\fi{}evi\ifmmode~\acute{c}\else \'{c}\fi{}},
  \citenamefont {Gjestvang}, \citenamefont {Canavan}, \citenamefont {Rudigier},
  \citenamefont {Étasse}, \citenamefont {Gerst}, \citenamefont {Gaudefroy},
  \citenamefont {Adamska}, \citenamefont {Adsley}, \citenamefont {Algora},
  \citenamefont {Babo}, \citenamefont {Belvedere}, \citenamefont {Benito},
  \citenamefont {Benzoni}, \citenamefont {Blazhev}, \citenamefont {Boso},
  \citenamefont {Bottoni}, \citenamefont {Bunce}, \citenamefont {Chakma},
  \citenamefont {Cieplicka-Oryńczak}, \citenamefont {Courtin}, \citenamefont
  {Cortés}, \citenamefont {Davies}, \citenamefont {Delafosse}, \citenamefont
  {Fallot}, \citenamefont {Fornal}, \citenamefont {Fraile}, \citenamefont
  {Gottardo}, \citenamefont {Guadilla}, \citenamefont {Häfner}, \citenamefont
  {Hauschild}, \citenamefont {Heine}, \citenamefont {Henrich}, \citenamefont
  {Homm}, \citenamefont {Ibrahim}, \citenamefont {Iskra}, \citenamefont
  {Ivanov}, \citenamefont {Jazrawi}, \citenamefont {Korgul}, \citenamefont
  {Koseoglou}, \citenamefont {Kröll}, \citenamefont {Kurtukian-Nieto},
  \citenamefont {Le~Meur}, \citenamefont {Leoni}, \citenamefont {Ljungvall},
  \citenamefont {Lopez-Martens}, \citenamefont {Lozeva}, \citenamefont {Matea},
  \citenamefont {Miernik}, \citenamefont {Nemer}, \citenamefont {Oberstedt},
  \citenamefont {Paulsen}, \citenamefont {Piersa}, \citenamefont {Popovitch},
  \citenamefont {Porzio}, \citenamefont {Qi}, \citenamefont {Ralet},
  \citenamefont {Regan}, \citenamefont {Rezynkina}, \citenamefont {S\ifmmode
  \acute{a}\else~\'{a}\fi{}nchez Tembleque}, \citenamefont {Siem},
  \citenamefont {Schmitt}, \citenamefont {Söderström}, \citenamefont
  {Sürder}, \citenamefont {Tocabens}, \citenamefont {Vedia}, \citenamefont
  {Verney}, \citenamefont {Warr}, \citenamefont {Wasilewska}, \citenamefont
  {Wiederhold}, \citenamefont {Yavahchova}, \citenamefont {Zeiser},\ and\
  \citenamefont {Ziliani}}]{Wilson2021Nature}%
  \BibitemOpen
  \bibfield  {author} {\bibinfo {author} {\bibfnamefont {J.~N.}\ \bibnamefont
  {Wilson}}, \bibinfo {author} {\bibfnamefont {D.}~\bibnamefont {Thisse}},
  \bibinfo {author} {\bibfnamefont {M.}~\bibnamefont {Lebois}}, \bibinfo
  {author} {\bibfnamefont {N.}~\bibnamefont {Jovan\ifmmode \check{c}\else
  \v{c}\fi{}evi\ifmmode~\acute{c}\else \'{c}\fi{}}}, \bibinfo {author}
  {\bibfnamefont {D.}~\bibnamefont {Gjestvang}}, \bibinfo {author}
  {\bibfnamefont {R.}~\bibnamefont {Canavan}}, \bibinfo {author} {\bibfnamefont
  {M.}~\bibnamefont {Rudigier}}, \bibinfo {author} {\bibfnamefont
  {D.}~\bibnamefont {Étasse}}, \bibinfo {author} {\bibfnamefont {R.-B.}\
  \bibnamefont {Gerst}}, \bibinfo {author} {\bibfnamefont {L.}~\bibnamefont
  {Gaudefroy}}, \bibinfo {author} {\bibfnamefont {E.}~\bibnamefont {Adamska}},
  \bibinfo {author} {\bibfnamefont {P.}~\bibnamefont {Adsley}}, \bibinfo
  {author} {\bibfnamefont {A.}~\bibnamefont {Algora}}, \bibinfo {author}
  {\bibfnamefont {M.}~\bibnamefont {Babo}}, \bibinfo {author} {\bibfnamefont
  {K.}~\bibnamefont {Belvedere}}, \bibinfo {author} {\bibfnamefont
  {J.}~\bibnamefont {Benito}}, \bibinfo {author} {\bibfnamefont
  {G.}~\bibnamefont {Benzoni}}, \bibinfo {author} {\bibfnamefont
  {A.}~\bibnamefont {Blazhev}}, \bibinfo {author} {\bibfnamefont
  {A.}~\bibnamefont {Boso}}, \bibinfo {author} {\bibfnamefont {S.}~\bibnamefont
  {Bottoni}}, \bibinfo {author} {\bibfnamefont {M.}~\bibnamefont {Bunce}},
  \bibinfo {author} {\bibfnamefont {R.}~\bibnamefont {Chakma}}, \bibinfo
  {author} {\bibfnamefont {N.}~\bibnamefont {Cieplicka-Oryńczak}}, \bibinfo
  {author} {\bibfnamefont {S.}~\bibnamefont {Courtin}}, \bibinfo {author}
  {\bibfnamefont {M.~L.}\ \bibnamefont {Cortés}}, \bibinfo {author}
  {\bibfnamefont {P.}~\bibnamefont {Davies}}, \bibinfo {author} {\bibfnamefont
  {C.}~\bibnamefont {Delafosse}}, \bibinfo {author} {\bibfnamefont
  {M.}~\bibnamefont {Fallot}}, \bibinfo {author} {\bibfnamefont
  {B.}~\bibnamefont {Fornal}}, \bibinfo {author} {\bibfnamefont
  {L.}~\bibnamefont {Fraile}}, \bibinfo {author} {\bibfnamefont
  {A.}~\bibnamefont {Gottardo}}, \bibinfo {author} {\bibfnamefont
  {V.}~\bibnamefont {Guadilla}}, \bibinfo {author} {\bibfnamefont
  {G.}~\bibnamefont {Häfner}}, \bibinfo {author} {\bibfnamefont
  {K.}~\bibnamefont {Hauschild}}, \bibinfo {author} {\bibfnamefont
  {M.}~\bibnamefont {Heine}}, \bibinfo {author} {\bibfnamefont
  {C.}~\bibnamefont {Henrich}}, \bibinfo {author} {\bibfnamefont
  {I.}~\bibnamefont {Homm}}, \bibinfo {author} {\bibfnamefont {F.}~\bibnamefont
  {Ibrahim}}, \bibinfo {author} {\bibfnamefont {.~W.}\ \bibnamefont {Iskra}},
  \bibinfo {author} {\bibfnamefont {P.}~\bibnamefont {Ivanov}}, \bibinfo
  {author} {\bibfnamefont {S.}~\bibnamefont {Jazrawi}}, \bibinfo {author}
  {\bibfnamefont {A.}~\bibnamefont {Korgul}}, \bibinfo {author} {\bibfnamefont
  {P.}~\bibnamefont {Koseoglou}}, \bibinfo {author} {\bibfnamefont
  {T.}~\bibnamefont {Kröll}}, \bibinfo {author} {\bibfnamefont
  {T.}~\bibnamefont {Kurtukian-Nieto}}, \bibinfo {author} {\bibfnamefont
  {L.}~\bibnamefont {Le~Meur}}, \bibinfo {author} {\bibfnamefont
  {S.}~\bibnamefont {Leoni}}, \bibinfo {author} {\bibfnamefont
  {J.}~\bibnamefont {Ljungvall}}, \bibinfo {author} {\bibfnamefont
  {A.}~\bibnamefont {Lopez-Martens}}, \bibinfo {author} {\bibfnamefont
  {R.}~\bibnamefont {Lozeva}}, \bibinfo {author} {\bibfnamefont
  {I.}~\bibnamefont {Matea}}, \bibinfo {author} {\bibfnamefont
  {K.}~\bibnamefont {Miernik}}, \bibinfo {author} {\bibfnamefont
  {J.}~\bibnamefont {Nemer}}, \bibinfo {author} {\bibfnamefont
  {S.}~\bibnamefont {Oberstedt}}, \bibinfo {author} {\bibfnamefont
  {W.}~\bibnamefont {Paulsen}}, \bibinfo {author} {\bibfnamefont
  {M.}~\bibnamefont {Piersa}}, \bibinfo {author} {\bibfnamefont
  {Y.}~\bibnamefont {Popovitch}}, \bibinfo {author} {\bibfnamefont
  {C.}~\bibnamefont {Porzio}}, \bibinfo {author} {\bibfnamefont
  {L.}~\bibnamefont {Qi}}, \bibinfo {author} {\bibfnamefont {D.}~\bibnamefont
  {Ralet}}, \bibinfo {author} {\bibfnamefont {P.~H.}\ \bibnamefont {Regan}},
  \bibinfo {author} {\bibfnamefont {K.}~\bibnamefont {Rezynkina}}, \bibinfo
  {author} {\bibfnamefont {V.}~\bibnamefont {S\ifmmode
  \acute{a}\else~\'{a}\fi{}nchez Tembleque}}, \bibinfo {author} {\bibfnamefont
  {S.}~\bibnamefont {Siem}}, \bibinfo {author} {\bibfnamefont {C.}~\bibnamefont
  {Schmitt}}, \bibinfo {author} {\bibfnamefont {P.-A.}\ \bibnamefont
  {Söderström}}, \bibinfo {author} {\bibfnamefont {C.}~\bibnamefont
  {Sürder}}, \bibinfo {author} {\bibfnamefont {G.}~\bibnamefont {Tocabens}},
  \bibinfo {author} {\bibfnamefont {V.}~\bibnamefont {Vedia}}, \bibinfo
  {author} {\bibfnamefont {D.}~\bibnamefont {Verney}}, \bibinfo {author}
  {\bibfnamefont {N.}~\bibnamefont {Warr}}, \bibinfo {author} {\bibfnamefont
  {B.}~\bibnamefont {Wasilewska}}, \bibinfo {author} {\bibfnamefont
  {J.}~\bibnamefont {Wiederhold}}, \bibinfo {author} {\bibfnamefont
  {M.}~\bibnamefont {Yavahchova}}, \bibinfo {author} {\bibfnamefont
  {F.}~\bibnamefont {Zeiser}}, \ and\ \bibinfo {author} {\bibfnamefont
  {S.}~\bibnamefont {Ziliani}},\ }\href {\doibase 10.1038/s41586-021-03304-w}
  {\bibfield  {journal} {\bibinfo  {journal} {Nature}\ }\textbf {\bibinfo
  {volume} {590}},\ \bibinfo {pages} {566} (\bibinfo {year}
  {2021})}\BibitemShut {NoStop}%
\bibitem [{\citenamefont {Marin}\ \emph {et~al.}(2021)\citenamefont {Marin},
  \citenamefont {Okar}, \citenamefont {Sansevero}, \citenamefont {Hernandez},
  \citenamefont {Ballard}, \citenamefont {Vogt}, \citenamefont {Randrup},
  \citenamefont {Talou}, \citenamefont {Lovell}, \citenamefont {Stetcu},
  \citenamefont {Serot}, \citenamefont {Litaize}, \citenamefont {Chebboubi},
  \citenamefont {Clarke}, \citenamefont {Protopopescu},\ and\ \citenamefont
  {Pozzi}}]{Marin2021PRC}%
  \BibitemOpen
  \bibfield  {author} {\bibinfo {author} {\bibfnamefont {S.}~\bibnamefont
  {Marin}}, \bibinfo {author} {\bibfnamefont {M.~S.}\ \bibnamefont {Okar}},
  \bibinfo {author} {\bibfnamefont {E.~P.}\ \bibnamefont {Sansevero}}, \bibinfo
  {author} {\bibfnamefont {I.~E.}\ \bibnamefont {Hernandez}}, \bibinfo {author}
  {\bibfnamefont {C.~A.}\ \bibnamefont {Ballard}}, \bibinfo {author}
  {\bibfnamefont {R.}~\bibnamefont {Vogt}}, \bibinfo {author} {\bibfnamefont
  {J.}~\bibnamefont {Randrup}}, \bibinfo {author} {\bibfnamefont
  {P.}~\bibnamefont {Talou}}, \bibinfo {author} {\bibfnamefont {A.~E.}\
  \bibnamefont {Lovell}}, \bibinfo {author} {\bibfnamefont {I.}~\bibnamefont
  {Stetcu}}, \bibinfo {author} {\bibfnamefont {O.}~\bibnamefont {Serot}},
  \bibinfo {author} {\bibfnamefont {O.}~\bibnamefont {Litaize}}, \bibinfo
  {author} {\bibfnamefont {A.}~\bibnamefont {Chebboubi}}, \bibinfo {author}
  {\bibfnamefont {S.~D.}\ \bibnamefont {Clarke}}, \bibinfo {author}
  {\bibfnamefont {V.~A.}\ \bibnamefont {Protopopescu}}, \ and\ \bibinfo
  {author} {\bibfnamefont {S.~A.}\ \bibnamefont {Pozzi}},\ }\href {\doibase
  10.1103/PhysRevC.104.024602} {\bibfield  {journal} {\bibinfo  {journal}
  {Phys. Rev. C}\ }\textbf {\bibinfo {volume} {104}},\ \bibinfo {pages}
  {024602} (\bibinfo {year} {2021})}\BibitemShut {NoStop}%
\bibitem [{\citenamefont {Marin}\ \emph {et~al.}(2022)\citenamefont {Marin},
  \citenamefont {Sansevero}, \citenamefont {Okar}, \citenamefont {Hernandez},
  \citenamefont {Vogt}, \citenamefont {Randrup}, \citenamefont {Clarke},
  \citenamefont {Protopopescu},\ and\ \citenamefont {Pozzi}}]{Marin2022PRC}%
  \BibitemOpen
  \bibfield  {author} {\bibinfo {author} {\bibfnamefont {S.}~\bibnamefont
  {Marin}}, \bibinfo {author} {\bibfnamefont {E.~P.}\ \bibnamefont
  {Sansevero}}, \bibinfo {author} {\bibfnamefont {M.~S.}\ \bibnamefont {Okar}},
  \bibinfo {author} {\bibfnamefont {I.~E.}\ \bibnamefont {Hernandez}}, \bibinfo
  {author} {\bibfnamefont {R.}~\bibnamefont {Vogt}}, \bibinfo {author}
  {\bibfnamefont {J.}~\bibnamefont {Randrup}}, \bibinfo {author} {\bibfnamefont
  {S.~D.}\ \bibnamefont {Clarke}}, \bibinfo {author} {\bibfnamefont {V.~A.}\
  \bibnamefont {Protopopescu}}, \ and\ \bibinfo {author} {\bibfnamefont
  {S.~A.}\ \bibnamefont {Pozzi}},\ }\href {\doibase
  10.1103/PhysRevC.105.054609} {\bibfield  {journal} {\bibinfo  {journal}
  {Phys. Rev. C}\ }\textbf {\bibinfo {volume} {105}},\ \bibinfo {pages}
  {054609} (\bibinfo {year} {2022})}\BibitemShut {NoStop}%
\bibitem [{\citenamefont {Giha}\ \emph {et~al.}(2023)\citenamefont {Giha},
  \citenamefont {Marin}, \citenamefont {Baker}, \citenamefont {Hernandez},
  \citenamefont {Kelly}, \citenamefont {Devlin}, \citenamefont {O'Donnell},
  \citenamefont {Vogt}, \citenamefont {Randrup}, \citenamefont {Talou},
  \citenamefont {Stetcu}, \citenamefont {Lovell}, \citenamefont {Litaize},
  \citenamefont {Serot}, \citenamefont {Chebboubi}, \citenamefont {Wu},
  \citenamefont {Clarke},\ and\ \citenamefont {Pozzi}}]{Giha2023PRC}%
  \BibitemOpen
  \bibfield  {author} {\bibinfo {author} {\bibfnamefont {N.~P.}\ \bibnamefont
  {Giha}}, \bibinfo {author} {\bibfnamefont {S.}~\bibnamefont {Marin}},
  \bibinfo {author} {\bibfnamefont {J.~A.}\ \bibnamefont {Baker}}, \bibinfo
  {author} {\bibfnamefont {I.~E.}\ \bibnamefont {Hernandez}}, \bibinfo {author}
  {\bibfnamefont {K.~J.}\ \bibnamefont {Kelly}}, \bibinfo {author}
  {\bibfnamefont {M.}~\bibnamefont {Devlin}}, \bibinfo {author} {\bibfnamefont
  {J.~M.}\ \bibnamefont {O'Donnell}}, \bibinfo {author} {\bibfnamefont
  {R.}~\bibnamefont {Vogt}}, \bibinfo {author} {\bibfnamefont {J.}~\bibnamefont
  {Randrup}}, \bibinfo {author} {\bibfnamefont {P.}~\bibnamefont {Talou}},
  \bibinfo {author} {\bibfnamefont {I.}~\bibnamefont {Stetcu}}, \bibinfo
  {author} {\bibfnamefont {A.~E.}\ \bibnamefont {Lovell}}, \bibinfo {author}
  {\bibfnamefont {O.}~\bibnamefont {Litaize}}, \bibinfo {author} {\bibfnamefont
  {O.}~\bibnamefont {Serot}}, \bibinfo {author} {\bibfnamefont
  {A.}~\bibnamefont {Chebboubi}}, \bibinfo {author} {\bibfnamefont {C.-Y.}\
  \bibnamefont {Wu}}, \bibinfo {author} {\bibfnamefont {S.~D.}\ \bibnamefont
  {Clarke}}, \ and\ \bibinfo {author} {\bibfnamefont {S.~A.}\ \bibnamefont
  {Pozzi}},\ }\href {\doibase 10.1103/PhysRevC.107.014612} {\bibfield
  {journal} {\bibinfo  {journal} {Phys. Rev. C}\ }\textbf {\bibinfo {volume}
  {107}},\ \bibinfo {pages} {014612} (\bibinfo {year} {2023})}\BibitemShut
  {NoStop}%
\bibitem [{\citenamefont {Marin}\ \emph {et~al.}(2024)\citenamefont {Marin},
  \citenamefont {Tolstukhin}, \citenamefont {Giha}, \citenamefont {Tovesson},
  \citenamefont {Protopopescu},\ and\ \citenamefont {Pozzi}}]{Marin2024PRC}%
  \BibitemOpen
  \bibfield  {author} {\bibinfo {author} {\bibfnamefont {S.}~\bibnamefont
  {Marin}}, \bibinfo {author} {\bibfnamefont {I.~A.}\ \bibnamefont
  {Tolstukhin}}, \bibinfo {author} {\bibfnamefont {N.~P.}\ \bibnamefont
  {Giha}}, \bibinfo {author} {\bibfnamefont {F.}~\bibnamefont {Tovesson}},
  \bibinfo {author} {\bibfnamefont {V.}~\bibnamefont {Protopopescu}}, \ and\
  \bibinfo {author} {\bibfnamefont {S.~A.}\ \bibnamefont {Pozzi}},\ }\href
  {\doibase 10.1103/PhysRevC.109.054617} {\bibfield  {journal} {\bibinfo
  {journal} {Phys. Rev. C}\ }\textbf {\bibinfo {volume} {109}},\ \bibinfo
  {pages} {054617} (\bibinfo {year} {2024})}\BibitemShut {NoStop}%
\bibitem [{\citenamefont {Bulgac}\ \emph {et~al.}(2021)\citenamefont {Bulgac},
  \citenamefont {Abdurrahman}, \citenamefont {Jin}, \citenamefont {Godbey},
  \citenamefont {Schunck},\ and\ \citenamefont {Stetcu}}]{Bulgac2021PRL}%
  \BibitemOpen
  \bibfield  {author} {\bibinfo {author} {\bibfnamefont {A.}~\bibnamefont
  {Bulgac}}, \bibinfo {author} {\bibfnamefont {I.}~\bibnamefont {Abdurrahman}},
  \bibinfo {author} {\bibfnamefont {S.}~\bibnamefont {Jin}}, \bibinfo {author}
  {\bibfnamefont {K.}~\bibnamefont {Godbey}}, \bibinfo {author} {\bibfnamefont
  {N.}~\bibnamefont {Schunck}}, \ and\ \bibinfo {author} {\bibfnamefont
  {I.}~\bibnamefont {Stetcu}},\ }\href {\doibase
  10.1103/PhysRevLett.126.142502} {\bibfield  {journal} {\bibinfo  {journal}
  {Phys. Rev. Lett.}\ }\textbf {\bibinfo {volume} {126}},\ \bibinfo {pages}
  {142502} (\bibinfo {year} {2021})}\BibitemShut {NoStop}%
\bibitem [{\citenamefont {Bulgac}\ \emph {et~al.}(2022)\citenamefont {Bulgac},
  \citenamefont {Abdurrahman}, \citenamefont {Godbey},\ and\ \citenamefont
  {Stetcu}}]{Bulgac2022PRL}%
  \BibitemOpen
  \bibfield  {author} {\bibinfo {author} {\bibfnamefont {A.}~\bibnamefont
  {Bulgac}}, \bibinfo {author} {\bibfnamefont {I.}~\bibnamefont {Abdurrahman}},
  \bibinfo {author} {\bibfnamefont {K.}~\bibnamefont {Godbey}}, \ and\ \bibinfo
  {author} {\bibfnamefont {I.}~\bibnamefont {Stetcu}},\ }\href {\doibase
  10.1103/PhysRevLett.128.022501} {\bibfield  {journal} {\bibinfo  {journal}
  {Phys. Rev. Lett.}\ }\textbf {\bibinfo {volume} {128}},\ \bibinfo {pages}
  {022501} (\bibinfo {year} {2022})}\BibitemShut {NoStop}%
\bibitem [{\citenamefont {Scamps}(2022)}]{Scamps2022PRC}%
  \BibitemOpen
  \bibfield  {author} {\bibinfo {author} {\bibfnamefont {G.}~\bibnamefont
  {Scamps}},\ }\href {\doibase 10.1103/PhysRevC.106.054614} {\bibfield
  {journal} {\bibinfo  {journal} {Phys. Rev. C}\ }\textbf {\bibinfo {volume}
  {106}},\ \bibinfo {pages} {054614} (\bibinfo {year} {2022})}\BibitemShut
  {NoStop}%
\bibitem [{\citenamefont {Scamps}\ \emph {et~al.}(2023)\citenamefont {Scamps},
  \citenamefont {Abdurrahman}, \citenamefont {Kafker}, \citenamefont {Bulgac},\
  and\ \citenamefont {Stetcu}}]{Scamps2023PRC}%
  \BibitemOpen
  \bibfield  {author} {\bibinfo {author} {\bibfnamefont {G.}~\bibnamefont
  {Scamps}}, \bibinfo {author} {\bibfnamefont {I.}~\bibnamefont {Abdurrahman}},
  \bibinfo {author} {\bibfnamefont {M.}~\bibnamefont {Kafker}}, \bibinfo
  {author} {\bibfnamefont {A.}~\bibnamefont {Bulgac}}, \ and\ \bibinfo {author}
  {\bibfnamefont {I.}~\bibnamefont {Stetcu}},\ }\href {\doibase
  10.1103/PhysRevC.108.L061602} {\bibfield  {journal} {\bibinfo  {journal}
  {Phys. Rev. C}\ }\textbf {\bibinfo {volume} {108}},\ \bibinfo {pages}
  {L061602} (\bibinfo {year} {2023})}\BibitemShut {NoStop}%
\bibitem [{\citenamefont {Marevi\ifmmode~\acute{c}\else \'{c}\fi{}}\ \emph
  {et~al.}(2021)\citenamefont {Marevi\ifmmode~\acute{c}\else \'{c}\fi{}},
  \citenamefont {Schunck}, \citenamefont {Randrup},\ and\ \citenamefont
  {Vogt}}]{Marevif2021PRC}%
  \BibitemOpen
  \bibfield  {author} {\bibinfo {author} {\bibfnamefont {P.}~\bibnamefont
  {Marevi\ifmmode~\acute{c}\else \'{c}\fi{}}}, \bibinfo {author} {\bibfnamefont
  {N.}~\bibnamefont {Schunck}}, \bibinfo {author} {\bibfnamefont
  {J.}~\bibnamefont {Randrup}}, \ and\ \bibinfo {author} {\bibfnamefont
  {R.}~\bibnamefont {Vogt}},\ }\href {\doibase 10.1103/PhysRevC.104.L021601}
  {\bibfield  {journal} {\bibinfo  {journal} {Phys. Rev. C}\ }\textbf {\bibinfo
  {volume} {104}},\ \bibinfo {pages} {L021601} (\bibinfo {year}
  {2021})}\BibitemShut {NoStop}%
\bibitem [{\citenamefont {Marevic}\ \emph {et~al.}()\citenamefont {Marevic},
  \citenamefont {Schunck},\ and\ \citenamefont {Verriere}}]{Marevic2025}%
  \BibitemOpen
  \bibfield  {author} {\bibinfo {author} {\bibfnamefont {P.}~\bibnamefont
  {Marevic}}, \bibinfo {author} {\bibfnamefont {N.}~\bibnamefont {Schunck}}, \
  and\ \bibinfo {author} {\bibfnamefont {M.}~\bibnamefont {Verriere}},\ }\href
  {https://arxiv.org/abs/2506.10777} {}\Eprint
  {http://arxiv.org/abs/2506.10777} {arXiv:2506.10777 [nucl-th]} \BibitemShut
  {NoStop}%
\bibitem [{\citenamefont {Randrup}\ and\ \citenamefont
  {Vogt}(2021)}]{Randrup2021PRL}%
  \BibitemOpen
  \bibfield  {author} {\bibinfo {author} {\bibfnamefont {J.}~\bibnamefont
  {Randrup}}\ and\ \bibinfo {author} {\bibfnamefont {R.}~\bibnamefont {Vogt}},\
  }\href {\doibase 10.1103/PhysRevLett.127.062502} {\bibfield  {journal}
  {\bibinfo  {journal} {Phys. Rev. Lett.}\ }\textbf {\bibinfo {volume} {127}},\
  \bibinfo {pages} {062502} (\bibinfo {year} {2021})}\BibitemShut {NoStop}%
\bibitem [{\citenamefont {Randrup}(2022)}]{Randrup2022PRC}%
  \BibitemOpen
  \bibfield  {author} {\bibinfo {author} {\bibfnamefont {J.}~\bibnamefont
  {Randrup}},\ }\href {\doibase 10.1103/PhysRevC.106.L051601} {\bibfield
  {journal} {\bibinfo  {journal} {Phys. Rev. C}\ }\textbf {\bibinfo {volume}
  {106}},\ \bibinfo {pages} {L051601} (\bibinfo {year} {2022})}\BibitemShut
  {NoStop}%
\bibitem [{\citenamefont {Randrup}\ \emph {et~al.}(2022)\citenamefont
  {Randrup}, \citenamefont {D\o{}ssing},\ and\ \citenamefont
  {Vogt}}]{Randrup2022PRC2}%
  \BibitemOpen
  \bibfield  {author} {\bibinfo {author} {\bibfnamefont {J.}~\bibnamefont
  {Randrup}}, \bibinfo {author} {\bibfnamefont {T.}~\bibnamefont {D\o{}ssing}},
  \ and\ \bibinfo {author} {\bibfnamefont {R.}~\bibnamefont {Vogt}},\ }\href
  {\doibase 10.1103/PhysRevC.106.014609} {\bibfield  {journal} {\bibinfo
  {journal} {Phys. Rev. C}\ }\textbf {\bibinfo {volume} {106}},\ \bibinfo
  {pages} {014609} (\bibinfo {year} {2022})}\BibitemShut {NoStop}%
\bibitem [{\citenamefont {Randrup}(2023)}]{Randrup2023PRC}%
  \BibitemOpen
  \bibfield  {author} {\bibinfo {author} {\bibfnamefont {J.}~\bibnamefont
  {Randrup}},\ }\href {\doibase 10.1103/PhysRevC.108.064606} {\bibfield
  {journal} {\bibinfo  {journal} {Phys. Rev. C}\ }\textbf {\bibinfo {volume}
  {108}},\ \bibinfo {pages} {064606} (\bibinfo {year} {2023})}\BibitemShut
  {NoStop}%
\bibitem [{\citenamefont {Zhou}\ \emph {et~al.}(2024)\citenamefont {Zhou},
  \citenamefont {Chen}, \citenamefont {Li}, \citenamefont {Smith},\ and\
  \citenamefont {Li}}]{Zhou2024}%
  \BibitemOpen
  \bibfield  {author} {\bibinfo {author} {\bibfnamefont {M.~H.}\ \bibnamefont
  {Zhou}}, \bibinfo {author} {\bibfnamefont {S.~Y.}\ \bibnamefont {Chen}},
  \bibinfo {author} {\bibfnamefont {Z.~Y.}\ \bibnamefont {Li}}, \bibinfo
  {author} {\bibfnamefont {M.~S.}\ \bibnamefont {Smith}}, \ and\ \bibinfo
  {author} {\bibfnamefont {Z.~P.}\ \bibnamefont {Li}},\ }\href
  {https://arxiv.org/abs/2311.06177} {\enquote {\bibinfo {title} {Quantum
  fluctuations drive angular momenta in nuclear fission},}\ } (\bibinfo {year}
  {2024}),\ \Eprint {http://arxiv.org/abs/2311.06177} {arXiv:2311.06177
  [nucl-th]} \BibitemShut {NoStop}%
\bibitem [{\citenamefont {D\o{}ssing}\ \emph {et~al.}(2024)\citenamefont
  {D\o{}ssing}, \citenamefont {\AA{}berg}, \citenamefont {Albertsson},
  \citenamefont {Carlsson},\ and\ \citenamefont {Randrup}}]{Dossing2024PRC}%
  \BibitemOpen
  \bibfield  {author} {\bibinfo {author} {\bibfnamefont {T.}~\bibnamefont
  {D\o{}ssing}}, \bibinfo {author} {\bibfnamefont {S.}~\bibnamefont
  {\AA{}berg}}, \bibinfo {author} {\bibfnamefont {M.}~\bibnamefont
  {Albertsson}}, \bibinfo {author} {\bibfnamefont {B.~G.}\ \bibnamefont
  {Carlsson}}, \ and\ \bibinfo {author} {\bibfnamefont {J.}~\bibnamefont
  {Randrup}},\ }\href {\doibase 10.1103/PhysRevC.109.034615} {\bibfield
  {journal} {\bibinfo  {journal} {Phys. Rev. C}\ }\textbf {\bibinfo {volume}
  {109}},\ \bibinfo {pages} {034615} (\bibinfo {year} {2024})}\BibitemShut
  {NoStop}%
\bibitem [{\citenamefont {Shneidman}\ \emph {et~al.}(2025)\citenamefont
  {Shneidman}, \citenamefont {Rahmatinejad}, \citenamefont {Adamian},\ and\
  \citenamefont {Antonenko}}]{Shneidman2025PRC}%
  \BibitemOpen
  \bibfield  {author} {\bibinfo {author} {\bibfnamefont {T.~M.}\ \bibnamefont
  {Shneidman}}, \bibinfo {author} {\bibfnamefont {A.}~\bibnamefont
  {Rahmatinejad}}, \bibinfo {author} {\bibfnamefont {G.~G.}\ \bibnamefont
  {Adamian}}, \ and\ \bibinfo {author} {\bibfnamefont {N.~V.}\ \bibnamefont
  {Antonenko}},\ }\href {\doibase 10.1103/ltv2-4141} {\bibfield  {journal}
  {\bibinfo  {journal} {Phys. Rev. C}\ }\textbf {\bibinfo {volume} {111}},\
  \bibinfo {pages} {064621} (\bibinfo {year} {2025})}\BibitemShut {NoStop}%
\bibitem [{\citenamefont {Scamps}\ and\ \citenamefont
  {Bertsch}(2023)}]{Scamps2023PRC2}%
  \BibitemOpen
  \bibfield  {author} {\bibinfo {author} {\bibfnamefont {G.}~\bibnamefont
  {Scamps}}\ and\ \bibinfo {author} {\bibfnamefont {G.}~\bibnamefont
  {Bertsch}},\ }\href {\doibase 10.1103/PhysRevC.108.034616} {\bibfield
  {journal} {\bibinfo  {journal} {Phys. Rev. C}\ }\textbf {\bibinfo {volume}
  {108}},\ \bibinfo {pages} {034616} (\bibinfo {year} {2023})}\BibitemShut
  {NoStop}%
\bibitem [{\citenamefont {Bertsch}\ \emph {et~al.}(2019)\citenamefont
  {Bertsch}, \citenamefont {Kawano},\ and\ \citenamefont
  {Robledo}}]{Bertsch2019PRC}%
  \BibitemOpen
  \bibfield  {author} {\bibinfo {author} {\bibfnamefont {G.~F.}\ \bibnamefont
  {Bertsch}}, \bibinfo {author} {\bibfnamefont {T.}~\bibnamefont {Kawano}}, \
  and\ \bibinfo {author} {\bibfnamefont {L.~M.}\ \bibnamefont {Robledo}},\
  }\href {\doibase 10.1103/PhysRevC.99.034603} {\bibfield  {journal} {\bibinfo
  {journal} {Phys. Rev. C}\ }\textbf {\bibinfo {volume} {99}},\ \bibinfo
  {pages} {034603} (\bibinfo {year} {2019})}\BibitemShut {NoStop}%
\bibitem [{\citenamefont {Bulgac}(2022)}]{Bulgac2022PRC}%
  \BibitemOpen
  \bibfield  {author} {\bibinfo {author} {\bibfnamefont {A.}~\bibnamefont
  {Bulgac}},\ }\href {\doibase 10.1103/PhysRevC.106.014624} {\bibfield
  {journal} {\bibinfo  {journal} {Phys. Rev. C}\ }\textbf {\bibinfo {volume}
  {106}},\ \bibinfo {pages} {014624} (\bibinfo {year} {2022})}\BibitemShut
  {NoStop}%
\bibitem [{\citenamefont {Scamps}(2024)}]{Scamps2024PRC}%
  \BibitemOpen
  \bibfield  {author} {\bibinfo {author} {\bibfnamefont {G.}~\bibnamefont
  {Scamps}},\ }\href {\doibase 10.1103/PhysRevC.109.L011602} {\bibfield
  {journal} {\bibinfo  {journal} {Phys. Rev. C}\ }\textbf {\bibinfo {volume}
  {109}},\ \bibinfo {pages} {L011602} (\bibinfo {year} {2024})}\BibitemShut
  {NoStop}%
\bibitem [{\citenamefont {Stetcu}\ \emph {et~al.}(2021)\citenamefont {Stetcu},
  \citenamefont {Lovell}, \citenamefont {Talou}, \citenamefont {Kawano},
  \citenamefont {Marin}, \citenamefont {Pozzi},\ and\ \citenamefont
  {Bulgac}}]{Stetcu2021PRL}%
  \BibitemOpen
  \bibfield  {author} {\bibinfo {author} {\bibfnamefont {I.}~\bibnamefont
  {Stetcu}}, \bibinfo {author} {\bibfnamefont {A.~E.}\ \bibnamefont {Lovell}},
  \bibinfo {author} {\bibfnamefont {P.}~\bibnamefont {Talou}}, \bibinfo
  {author} {\bibfnamefont {T.}~\bibnamefont {Kawano}}, \bibinfo {author}
  {\bibfnamefont {S.}~\bibnamefont {Marin}}, \bibinfo {author} {\bibfnamefont
  {S.~A.}\ \bibnamefont {Pozzi}}, \ and\ \bibinfo {author} {\bibfnamefont
  {A.}~\bibnamefont {Bulgac}},\ }\href {\doibase
  10.1103/PhysRevLett.127.222502} {\bibfield  {journal} {\bibinfo  {journal}
  {Phys. Rev. Lett.}\ }\textbf {\bibinfo {volume} {127}},\ \bibinfo {pages}
  {222502} (\bibinfo {year} {2021})}\BibitemShut {NoStop}%
\bibitem [{\citenamefont {Bohr}\ and\ \citenamefont
  {Mottelson}(1975)}]{Bohr1975}%
  \BibitemOpen
  \bibfield  {author} {\bibinfo {author} {\bibfnamefont {A.}~\bibnamefont
  {Bohr}}\ and\ \bibinfo {author} {\bibfnamefont {B.}~\bibnamefont
  {Mottelson}},\ }\href {https://api.semanticscholar.org/CorpusID:117271202}
  {\emph {\bibinfo {title} {Nuclear Structure, Volume II: Nuclear
  Deformations}}}\ (\bibinfo {year} {1975})\BibitemShut {NoStop}%
\bibitem [{\citenamefont {Ring}\ and\ \citenamefont
  {Schuck}(2004)}]{RingManybody}%
  \BibitemOpen
  \bibfield  {author} {\bibinfo {author} {\bibfnamefont {P.}~\bibnamefont
  {Ring}}\ and\ \bibinfo {author} {\bibfnamefont {P.}~\bibnamefont {Schuck}},\
  }\href@noop {} {\emph {\bibinfo {title} {The Nuclear Many-Body Problem}}}\
  (\bibinfo  {publisher} {Springer Science and Business Media, New York},\
  \bibinfo {year} {2004})\BibitemShut {NoStop}%
\bibitem [{\citenamefont {Meng}(2016)}]{Meng2016}%
  \BibitemOpen
  \bibfield  {author} {\bibinfo {author} {\bibfnamefont {J.}~\bibnamefont
  {Meng}},\ }\href {\doibase 10.1142/9872} {\emph {\bibinfo {title}
  {Relativistic Density Functional for Nuclear Structure}}}\ (\bibinfo
  {publisher} {WORLD SCIENTIFIC},\ \bibinfo {year} {2016})\ \Eprint
  {http://arxiv.org/abs/https://www.worldscientific.com/doi/pdf/10.1142/9872}
  {https://www.worldscientific.com/doi/pdf/10.1142/9872} \BibitemShut {NoStop}%
\bibitem [{\citenamefont {Frauendorf}\ and\ \citenamefont {{Jie
  Meng}}(1997)}]{Frauendorf1997NPA}%
  \BibitemOpen
  \bibfield  {author} {\bibinfo {author} {\bibfnamefont {S.}~\bibnamefont
  {Frauendorf}}\ and\ \bibinfo {author} {\bibnamefont {{Jie Meng}}},\ }\href
  {\doibase https://doi.org/10.1016/S0375-9474(97)00004-3} {\bibfield
  {journal} {\bibinfo  {journal} {Nucl. Phys. A}\ }\textbf {\bibinfo {volume}
  {617}},\ \bibinfo {pages} {131} (\bibinfo {year} {1997})}\BibitemShut
  {NoStop}%
\bibitem [{\citenamefont {Lu}\ \emph {et~al.}(2012)\citenamefont {Lu},
  \citenamefont {Zhao},\ and\ \citenamefont {Zhou}}]{Lu2012PRC}%
  \BibitemOpen
  \bibfield  {author} {\bibinfo {author} {\bibfnamefont {B.-N.}\ \bibnamefont
  {Lu}}, \bibinfo {author} {\bibfnamefont {E.-G.}\ \bibnamefont {Zhao}}, \ and\
  \bibinfo {author} {\bibfnamefont {S.-G.}\ \bibnamefont {Zhou}},\ }\href
  {\doibase 10.1103/PhysRevC.85.011301} {\bibfield  {journal} {\bibinfo
  {journal} {Phys. Rev. C}\ }\textbf {\bibinfo {volume} {85}},\ \bibinfo
  {pages} {011301} (\bibinfo {year} {2012})}\BibitemShut {NoStop}%
\bibitem [{\citenamefont {Lu}\ \emph {et~al.}(2014)\citenamefont {Lu},
  \citenamefont {Zhao}, \citenamefont {Zhao},\ and\ \citenamefont
  {Zhou}}]{Lu2014PRC}%
  \BibitemOpen
  \bibfield  {author} {\bibinfo {author} {\bibfnamefont {B.-N.}\ \bibnamefont
  {Lu}}, \bibinfo {author} {\bibfnamefont {J.}~\bibnamefont {Zhao}}, \bibinfo
  {author} {\bibfnamefont {E.-G.}\ \bibnamefont {Zhao}}, \ and\ \bibinfo
  {author} {\bibfnamefont {S.-G.}\ \bibnamefont {Zhou}},\ }\href {\doibase
  10.1103/PhysRevC.89.014323} {\bibfield  {journal} {\bibinfo  {journal} {Phys.
  Rev. C}\ }\textbf {\bibinfo {volume} {89}},\ \bibinfo {pages} {014323}
  (\bibinfo {year} {2014})}\BibitemShut {NoStop}%
\bibitem [{\citenamefont {Ryssens}\ \emph {et~al.}(2023)\citenamefont
  {Ryssens}, \citenamefont {Scamps}, \citenamefont {Goriely},\ and\
  \citenamefont {Bender}}]{Ryssens2023EPJA}%
  \BibitemOpen
  \bibfield  {author} {\bibinfo {author} {\bibfnamefont {W.}~\bibnamefont
  {Ryssens}}, \bibinfo {author} {\bibfnamefont {G.}~\bibnamefont {Scamps}},
  \bibinfo {author} {\bibfnamefont {S.}~\bibnamefont {Goriely}}, \ and\
  \bibinfo {author} {\bibfnamefont {M.}~\bibnamefont {Bender}},\ }\href
  {\doibase 10.1140/epja/s10050-023-01002-x} {\bibfield  {journal} {\bibinfo
  {journal} {Eur. Phys. J. A}\ }\textbf {\bibinfo {volume} {59}},\ \bibinfo
  {pages} {96} (\bibinfo {year} {2023})}\BibitemShut {NoStop}%
\bibitem [{\citenamefont {Ebata}\ \emph {et~al.}(2010)\citenamefont {Ebata},
  \citenamefont {Nakatsukasa}, \citenamefont {Inakura}, \citenamefont
  {Yoshida}, \citenamefont {Hashimoto},\ and\ \citenamefont
  {Yabana}}]{Ebata2010PRC}%
  \BibitemOpen
  \bibfield  {author} {\bibinfo {author} {\bibfnamefont {S.}~\bibnamefont
  {Ebata}}, \bibinfo {author} {\bibfnamefont {T.}~\bibnamefont {Nakatsukasa}},
  \bibinfo {author} {\bibfnamefont {T.}~\bibnamefont {Inakura}}, \bibinfo
  {author} {\bibfnamefont {K.}~\bibnamefont {Yoshida}}, \bibinfo {author}
  {\bibfnamefont {Y.}~\bibnamefont {Hashimoto}}, \ and\ \bibinfo {author}
  {\bibfnamefont {K.}~\bibnamefont {Yabana}},\ }\href {\doibase
  10.1103/PhysRevC.82.034306} {\bibfield  {journal} {\bibinfo  {journal} {Phys.
  Rev. C}\ }\textbf {\bibinfo {volume} {82}},\ \bibinfo {pages} {034306}
  (\bibinfo {year} {2010})}\BibitemShut {NoStop}%
\bibitem [{\citenamefont {Scamps}\ and\ \citenamefont
  {Lacroix}(2013)}]{Scamps2013PRC}%
  \BibitemOpen
  \bibfield  {author} {\bibinfo {author} {\bibfnamefont {G.}~\bibnamefont
  {Scamps}}\ and\ \bibinfo {author} {\bibfnamefont {D.}~\bibnamefont
  {Lacroix}},\ }\href {\doibase 10.1103/PhysRevC.87.014605} {\bibfield
  {journal} {\bibinfo  {journal} {Phys. Rev. C}\ }\textbf {\bibinfo {volume}
  {87}},\ \bibinfo {pages} {014605} (\bibinfo {year} {2013})}\BibitemShut
  {NoStop}%
\bibitem [{\citenamefont {Magierski}\ \emph {et~al.}(2018)\citenamefont
  {Magierski}, \citenamefont {Grineviciute},\ and\ \citenamefont
  {Sekizawa}}]{Magierski2018APP}%
  \BibitemOpen
  \bibfield  {author} {\bibinfo {author} {\bibfnamefont {P.}~\bibnamefont
  {Magierski}}, \bibinfo {author} {\bibfnamefont {J.}~\bibnamefont
  {Grineviciute}}, \ and\ \bibinfo {author} {\bibfnamefont {K.}~\bibnamefont
  {Sekizawa}},\ }\href {\doibase 10.5506/aphyspolb.49.281} {\bibfield
  {journal} {\bibinfo  {journal} {Acta Physica Polonica B}\ }\textbf {\bibinfo
  {volume} {49}},\ \bibinfo {pages} {281} (\bibinfo {year} {2018})}\BibitemShut
  {NoStop}%
\bibitem [{\citenamefont {Scamps}\ \emph {et~al.}(2012)\citenamefont {Scamps},
  \citenamefont {Lacroix}, \citenamefont {Bertsch},\ and\ \citenamefont
  {Washiyama}}]{Scamps2012PRC}%
  \BibitemOpen
  \bibfield  {author} {\bibinfo {author} {\bibfnamefont {G.}~\bibnamefont
  {Scamps}}, \bibinfo {author} {\bibfnamefont {D.}~\bibnamefont {Lacroix}},
  \bibinfo {author} {\bibfnamefont {G.~F.}\ \bibnamefont {Bertsch}}, \ and\
  \bibinfo {author} {\bibfnamefont {K.}~\bibnamefont {Washiyama}},\ }\href
  {\doibase 10.1103/PhysRevC.85.034328} {\bibfield  {journal} {\bibinfo
  {journal} {Phys. Rev. C}\ }\textbf {\bibinfo {volume} {85}},\ \bibinfo
  {pages} {034328} (\bibinfo {year} {2012})}\BibitemShut {NoStop}%
\bibitem [{\citenamefont {Bulgac}(2007)}]{Bulgac2007PRA}%
  \BibitemOpen
  \bibfield  {author} {\bibinfo {author} {\bibfnamefont {A.}~\bibnamefont
  {Bulgac}},\ }\href {\doibase 10.1103/PhysRevA.76.040502} {\bibfield
  {journal} {\bibinfo  {journal} {Phys. Rev. A}\ }\textbf {\bibinfo {volume}
  {76}},\ \bibinfo {pages} {040502} (\bibinfo {year} {2007})}\BibitemShut
  {NoStop}%
\bibitem [{\citenamefont {Ren}\ \emph {et~al.}(2020{\natexlab{a}})\citenamefont
  {Ren}, \citenamefont {Zhao},\ and\ \citenamefont {Meng}}]{Ren2020PRC}%
  \BibitemOpen
  \bibfield  {author} {\bibinfo {author} {\bibfnamefont {Z.~X.}\ \bibnamefont
  {Ren}}, \bibinfo {author} {\bibfnamefont {P.~W.}\ \bibnamefont {Zhao}}, \
  and\ \bibinfo {author} {\bibfnamefont {J.}~\bibnamefont {Meng}},\ }\href
  {\doibase 10.1103/PhysRevC.102.044603} {\bibfield  {journal} {\bibinfo
  {journal} {Phys. Rev. C}\ }\textbf {\bibinfo {volume} {102}},\ \bibinfo
  {pages} {044603} (\bibinfo {year} {2020}{\natexlab{a}})}\BibitemShut
  {NoStop}%
\bibitem [{\citenamefont {Ren}\ \emph {et~al.}(2020{\natexlab{b}})\citenamefont
  {Ren}, \citenamefont {Zhao},\ and\ \citenamefont {Meng}}]{Ren2020PLB}%
  \BibitemOpen
  \bibfield  {author} {\bibinfo {author} {\bibfnamefont {Z.~X.}\ \bibnamefont
  {Ren}}, \bibinfo {author} {\bibfnamefont {P.~W.}\ \bibnamefont {Zhao}}, \
  and\ \bibinfo {author} {\bibfnamefont {J.}~\bibnamefont {Meng}},\ }\href
  {\doibase https://doi.org/10.1016/j.physletb.2019.135194} {\bibfield
  {journal} {\bibinfo  {journal} {Phys. Lett. B}\ }\textbf {\bibinfo {volume}
  {801}},\ \bibinfo {pages} {135194} (\bibinfo {year}
  {2020}{\natexlab{b}})}\BibitemShut {NoStop}%
\bibitem [{\citenamefont {Ren}\ \emph {et~al.}(2022{\natexlab{a}})\citenamefont
  {Ren}, \citenamefont {Vretenar}, \citenamefont {Nik\ifmmode \check{s}\else
  \v{s}\fi{}i\ifmmode~\acute{c}\else \'{c}\fi{}}, \citenamefont {Zhao},
  \citenamefont {Zhao},\ and\ \citenamefont {Meng}}]{Ren2022PRL}%
  \BibitemOpen
  \bibfield  {author} {\bibinfo {author} {\bibfnamefont {Z.~X.}\ \bibnamefont
  {Ren}}, \bibinfo {author} {\bibfnamefont {D.}~\bibnamefont {Vretenar}},
  \bibinfo {author} {\bibfnamefont {T.}~\bibnamefont {Nik\ifmmode
  \check{s}\else \v{s}\fi{}i\ifmmode~\acute{c}\else \'{c}\fi{}}}, \bibinfo
  {author} {\bibfnamefont {P.~W.}\ \bibnamefont {Zhao}}, \bibinfo {author}
  {\bibfnamefont {J.}~\bibnamefont {Zhao}}, \ and\ \bibinfo {author}
  {\bibfnamefont {J.}~\bibnamefont {Meng}},\ }\href {\doibase
  10.1103/PhysRevLett.128.172501} {\bibfield  {journal} {\bibinfo  {journal}
  {Phys. Rev. Lett.}\ }\textbf {\bibinfo {volume} {128}},\ \bibinfo {pages}
  {172501} (\bibinfo {year} {2022}{\natexlab{a}})}\BibitemShut {NoStop}%
\bibitem [{\citenamefont {Ren}\ \emph {et~al.}(2022{\natexlab{b}})\citenamefont
  {Ren}, \citenamefont {Zhao}, \citenamefont {Vretenar}, \citenamefont
  {Nik\ifmmode \check{s}\else \v{s}\fi{}i\ifmmode~\acute{c}\else \'{c}\fi{}},
  \citenamefont {Zhao},\ and\ \citenamefont {Meng}}]{Ren2022a}%
  \BibitemOpen
  \bibfield  {author} {\bibinfo {author} {\bibfnamefont {Z.~X.}\ \bibnamefont
  {Ren}}, \bibinfo {author} {\bibfnamefont {J.}~\bibnamefont {Zhao}}, \bibinfo
  {author} {\bibfnamefont {D.}~\bibnamefont {Vretenar}}, \bibinfo {author}
  {\bibfnamefont {T.}~\bibnamefont {Nik\ifmmode \check{s}\else
  \v{s}\fi{}i\ifmmode~\acute{c}\else \'{c}\fi{}}}, \bibinfo {author}
  {\bibfnamefont {P.~W.}\ \bibnamefont {Zhao}}, \ and\ \bibinfo {author}
  {\bibfnamefont {J.}~\bibnamefont {Meng}},\ }\href {\doibase
  10.1103/PhysRevC.105.044313} {\bibfield  {journal} {\bibinfo  {journal}
  {Phys. Rev. C}\ }\textbf {\bibinfo {volume} {105}},\ \bibinfo {pages}
  {044313} (\bibinfo {year} {2022}{\natexlab{b}})}\BibitemShut {NoStop}%
\bibitem [{\citenamefont {Ren}\ \emph {et~al.}(2022{\natexlab{c}})\citenamefont
  {Ren}, \citenamefont {Zhao},\ and\ \citenamefont {Meng}}]{Ren2022b}%
  \BibitemOpen
  \bibfield  {author} {\bibinfo {author} {\bibfnamefont {Z.~X.}\ \bibnamefont
  {Ren}}, \bibinfo {author} {\bibfnamefont {P.~W.}\ \bibnamefont {Zhao}}, \
  and\ \bibinfo {author} {\bibfnamefont {J.}~\bibnamefont {Meng}},\ }\href
  {\doibase 10.1103/PhysRevC.105.L011301} {\bibfield  {journal} {\bibinfo
  {journal} {Phys. Rev. C}\ }\textbf {\bibinfo {volume} {105}},\ \bibinfo
  {pages} {L011301} (\bibinfo {year} {2022}{\natexlab{c}})}\BibitemShut
  {NoStop}%
\bibitem [{\citenamefont {Zhang}\ \emph
  {et~al.}(2024{\natexlab{a}})\citenamefont {Zhang}, \citenamefont {Li},
  \citenamefont {Vretenar}, \citenamefont {Nik\ifmmode \check{s}\else
  \v{s}\fi{}i\ifmmode~\acute{c}\else \'{c}\fi{}}, \citenamefont {Ren},
  \citenamefont {Zhao},\ and\ \citenamefont {Meng}}]{Zhang2024PRCa}%
  \BibitemOpen
  \bibfield  {author} {\bibinfo {author} {\bibfnamefont {D.~D.}\ \bibnamefont
  {Zhang}}, \bibinfo {author} {\bibfnamefont {B.}~\bibnamefont {Li}}, \bibinfo
  {author} {\bibfnamefont {D.}~\bibnamefont {Vretenar}}, \bibinfo {author}
  {\bibfnamefont {T.}~\bibnamefont {Nik\ifmmode \check{s}\else
  \v{s}\fi{}i\ifmmode~\acute{c}\else \'{c}\fi{}}}, \bibinfo {author}
  {\bibfnamefont {Z.~X.}\ \bibnamefont {Ren}}, \bibinfo {author} {\bibfnamefont
  {P.~W.}\ \bibnamefont {Zhao}}, \ and\ \bibinfo {author} {\bibfnamefont
  {J.}~\bibnamefont {Meng}},\ }\href {\doibase 10.1103/PhysRevC.109.024316}
  {\bibfield  {journal} {\bibinfo  {journal} {Phys. Rev. C}\ }\textbf {\bibinfo
  {volume} {109}},\ \bibinfo {pages} {024316} (\bibinfo {year}
  {2024}{\natexlab{a}})}\BibitemShut {NoStop}%
\bibitem [{\citenamefont {Zhang}\ \emph
  {et~al.}(2024{\natexlab{b}})\citenamefont {Zhang}, \citenamefont {Vretenar},
  \citenamefont {Nik\ifmmode \check{s}\else \v{s}\fi{}i\ifmmode~\acute{c}\else
  \'{c}\fi{}}, \citenamefont {Zhao},\ and\ \citenamefont
  {Meng}}]{Zhang2024PRCb}%
  \BibitemOpen
  \bibfield  {author} {\bibinfo {author} {\bibfnamefont {D.~D.}\ \bibnamefont
  {Zhang}}, \bibinfo {author} {\bibfnamefont {D.}~\bibnamefont {Vretenar}},
  \bibinfo {author} {\bibfnamefont {T.}~\bibnamefont {Nik\ifmmode
  \check{s}\else \v{s}\fi{}i\ifmmode~\acute{c}\else \'{c}\fi{}}}, \bibinfo
  {author} {\bibfnamefont {P.~W.}\ \bibnamefont {Zhao}}, \ and\ \bibinfo
  {author} {\bibfnamefont {J.}~\bibnamefont {Meng}},\ }\href {\doibase
  10.1103/PhysRevC.109.024614} {\bibfield  {journal} {\bibinfo  {journal}
  {Phys. Rev. C}\ }\textbf {\bibinfo {volume} {109}},\ \bibinfo {pages}
  {024614} (\bibinfo {year} {2024}{\natexlab{b}})}\BibitemShut {NoStop}%
\bibitem [{\citenamefont {Li}\ \emph {et~al.}(2023{\natexlab{a}})\citenamefont
  {Li}, \citenamefont {Vretenar}, \citenamefont {Ren}, \citenamefont
  {Nik\ifmmode \check{s}\else \v{s}\fi{}i\ifmmode~\acute{c}\else \'{c}\fi{}},
  \citenamefont {Zhao}, \citenamefont {Zhao},\ and\ \citenamefont
  {Meng}}]{Li2023PRC_FTTD}%
  \BibitemOpen
  \bibfield  {author} {\bibinfo {author} {\bibfnamefont {B.}~\bibnamefont
  {Li}}, \bibinfo {author} {\bibfnamefont {D.}~\bibnamefont {Vretenar}},
  \bibinfo {author} {\bibfnamefont {Z.~X.}\ \bibnamefont {Ren}}, \bibinfo
  {author} {\bibfnamefont {T.}~\bibnamefont {Nik\ifmmode \check{s}\else
  \v{s}\fi{}i\ifmmode~\acute{c}\else \'{c}\fi{}}}, \bibinfo {author}
  {\bibfnamefont {J.}~\bibnamefont {Zhao}}, \bibinfo {author} {\bibfnamefont
  {P.~W.}\ \bibnamefont {Zhao}}, \ and\ \bibinfo {author} {\bibfnamefont
  {J.}~\bibnamefont {Meng}},\ }\href {\doibase 10.1103/PhysRevC.107.014303}
  {\bibfield  {journal} {\bibinfo  {journal} {Phys. Rev. C}\ }\textbf {\bibinfo
  {volume} {107}},\ \bibinfo {pages} {014303} (\bibinfo {year}
  {2023}{\natexlab{a}})}\BibitemShut {NoStop}%
\bibitem [{\citenamefont {Li}\ \emph {et~al.}(2024{\natexlab{a}})\citenamefont
  {Li}, \citenamefont {Vretenar}, \citenamefont {Nik\ifmmode \check{s}\else
  \v{s}\fi{}i\ifmmode~\acute{c}\else \'{c}\fi{}}, \citenamefont {Zhang},
  \citenamefont {Zhao},\ and\ \citenamefont {Meng}}]{Li2024PRCa}%
  \BibitemOpen
  \bibfield  {author} {\bibinfo {author} {\bibfnamefont {B.}~\bibnamefont
  {Li}}, \bibinfo {author} {\bibfnamefont {D.}~\bibnamefont {Vretenar}},
  \bibinfo {author} {\bibfnamefont {T.}~\bibnamefont {Nik\ifmmode
  \check{s}\else \v{s}\fi{}i\ifmmode~\acute{c}\else \'{c}\fi{}}}, \bibinfo
  {author} {\bibfnamefont {D.~D.}\ \bibnamefont {Zhang}}, \bibinfo {author}
  {\bibfnamefont {P.~W.}\ \bibnamefont {Zhao}}, \ and\ \bibinfo {author}
  {\bibfnamefont {J.}~\bibnamefont {Meng}},\ }\href {\doibase
  10.1103/PhysRevC.110.034611} {\bibfield  {journal} {\bibinfo  {journal}
  {Phys. Rev. C}\ }\textbf {\bibinfo {volume} {110}},\ \bibinfo {pages}
  {034611} (\bibinfo {year} {2024}{\natexlab{a}})}\BibitemShut {NoStop}%
\bibitem [{\citenamefont {Li}\ \emph {et~al.}(2024{\natexlab{b}})\citenamefont
  {Li}, \citenamefont {Vretenar}, \citenamefont {Nik\ifmmode \check{s}\else
  \v{s}\fi{}i\ifmmode~\acute{c}\else \'{c}\fi{}}, \citenamefont {Zhao},\ and\
  \citenamefont {Meng}}]{Li2024PRCb}%
  \BibitemOpen
  \bibfield  {author} {\bibinfo {author} {\bibfnamefont {B.}~\bibnamefont
  {Li}}, \bibinfo {author} {\bibfnamefont {D.}~\bibnamefont {Vretenar}},
  \bibinfo {author} {\bibfnamefont {T.}~\bibnamefont {Nik\ifmmode
  \check{s}\else \v{s}\fi{}i\ifmmode~\acute{c}\else \'{c}\fi{}}}, \bibinfo
  {author} {\bibfnamefont {P.~W.}\ \bibnamefont {Zhao}}, \ and\ \bibinfo
  {author} {\bibfnamefont {J.}~\bibnamefont {Meng}},\ }\href {\doibase
  10.1103/PhysRevC.110.034302} {\bibfield  {journal} {\bibinfo  {journal}
  {Phys. Rev. C}\ }\textbf {\bibinfo {volume} {110}},\ \bibinfo {pages}
  {034302} (\bibinfo {year} {2024}{\natexlab{b}})}\BibitemShut {NoStop}%
\bibitem [{\citenamefont {Li}\ \emph {et~al.}(2024{\natexlab{c}})\citenamefont
  {Li}, \citenamefont {Zhao},\ and\ \citenamefont {Meng}}]{Li2024PLB}%
  \BibitemOpen
  \bibfield  {author} {\bibinfo {author} {\bibfnamefont {B.}~\bibnamefont
  {Li}}, \bibinfo {author} {\bibfnamefont {P.~W.}\ \bibnamefont {Zhao}}, \ and\
  \bibinfo {author} {\bibfnamefont {J.}~\bibnamefont {Meng}},\ }\href {\doibase
  https://doi.org/10.1016/j.physletb.2024.138877} {\bibfield  {journal}
  {\bibinfo  {journal} {Phys. Lett. B}\ }\textbf {\bibinfo {volume} {856}},\
  \bibinfo {pages} {138877} (\bibinfo {year} {2024}{\natexlab{c}})}\BibitemShut
  {NoStop}%
\bibitem [{\citenamefont {Zhang}\ \emph {et~al.}(2025)\citenamefont {Zhang},
  \citenamefont {Vretenar}, \citenamefont {Nik\ifmmode \check{s}\else
  \v{s}\fi{}i\ifmmode~\acute{c}\else \'{c}\fi{}}, \citenamefont {Zhao},\ and\
  \citenamefont {Meng}}]{Zhang2025PLB}%
  \BibitemOpen
  \bibfield  {author} {\bibinfo {author} {\bibfnamefont {D.~D.}\ \bibnamefont
  {Zhang}}, \bibinfo {author} {\bibfnamefont {D.}~\bibnamefont {Vretenar}},
  \bibinfo {author} {\bibfnamefont {T.}~\bibnamefont {Nik\ifmmode
  \check{s}\else \v{s}\fi{}i\ifmmode~\acute{c}\else \'{c}\fi{}}}, \bibinfo
  {author} {\bibfnamefont {P.~W.}\ \bibnamefont {Zhao}}, \ and\ \bibinfo
  {author} {\bibfnamefont {J.}~\bibnamefont {Meng}},\ }\href {\doibase
  https://doi.org/10.1016/j.physletb.2025.139828} {\bibfield  {journal}
  {\bibinfo  {journal} {Phys. Lett. B}\ }\textbf {\bibinfo {volume} {869}},\
  \bibinfo {pages} {139828} (\bibinfo {year} {2025})}\BibitemShut {NoStop}%
\bibitem [{\citenamefont {Li}\ \emph {et~al.}()\citenamefont {Li},
  \citenamefont {Zhang}, \citenamefont {Vretenar}, \citenamefont {Nik\ifmmode
  \check{s}\else \v{s}\fi{}i\ifmmode~\acute{c}\else \'{c}\fi{}}, \citenamefont
  {Zhao},\ and\ \citenamefont {Meng}}]{Li2025Supp}%
  \BibitemOpen
  \bibfield  {author} {\bibinfo {author} {\bibfnamefont {B.}~\bibnamefont
  {Li}}, \bibinfo {author} {\bibfnamefont {D.~D.}\ \bibnamefont {Zhang}},
  \bibinfo {author} {\bibfnamefont {D.}~\bibnamefont {Vretenar}}, \bibinfo
  {author} {\bibfnamefont {T.}~\bibnamefont {Nik\ifmmode \check{s}\else
  \v{s}\fi{}i\ifmmode~\acute{c}\else \'{c}\fi{}}}, \bibinfo {author}
  {\bibfnamefont {P.~W.}\ \bibnamefont {Zhao}}, \ and\ \bibinfo {author}
  {\bibfnamefont {J.}~\bibnamefont {Meng}},\ }\href@noop {} {\enquote {\bibinfo
  {title} {Supplemental material to this article at [url], which includes
  additional Ref. \cite{Hu2014PLB}.}}\ }\BibitemShut {NoStop}%
\bibitem [{\citenamefont {Ren}\ \emph {et~al.}(2017)\citenamefont {Ren},
  \citenamefont {Zhang},\ and\ \citenamefont {Meng}}]{Ren2017PRC}%
  \BibitemOpen
  \bibfield  {author} {\bibinfo {author} {\bibfnamefont {Z.~X.}\ \bibnamefont
  {Ren}}, \bibinfo {author} {\bibfnamefont {S.~Q.}\ \bibnamefont {Zhang}}, \
  and\ \bibinfo {author} {\bibfnamefont {J.}~\bibnamefont {Meng}},\ }\href
  {\doibase 10.1103/PhysRevC.95.024313} {\bibfield  {journal} {\bibinfo
  {journal} {Phys. Rev. C}\ }\textbf {\bibinfo {volume} {95}},\ \bibinfo
  {pages} {024313} (\bibinfo {year} {2017})}\BibitemShut {NoStop}%
\bibitem [{\citenamefont {Ren}\ \emph {et~al.}(2019)\citenamefont {Ren},
  \citenamefont {Zhang}, \citenamefont {Zhao}, \citenamefont {Itagaki},
  \citenamefont {Maruhn},\ and\ \citenamefont {Meng}}]{Ren2019SCPMA}%
  \BibitemOpen
  \bibfield  {author} {\bibinfo {author} {\bibfnamefont {Z.~X.}\ \bibnamefont
  {Ren}}, \bibinfo {author} {\bibfnamefont {S.~Q.}\ \bibnamefont {Zhang}},
  \bibinfo {author} {\bibfnamefont {P.~W.}\ \bibnamefont {Zhao}}, \bibinfo
  {author} {\bibfnamefont {N.}~\bibnamefont {Itagaki}}, \bibinfo {author}
  {\bibfnamefont {J.~A.}\ \bibnamefont {Maruhn}}, \ and\ \bibinfo {author}
  {\bibfnamefont {J.}~\bibnamefont {Meng}},\ }\href {\doibase
  10.1007/s11433-019-9412-3} {\bibfield  {journal} {\bibinfo  {journal} {Sci.
  China Phys. Mech. Astron.}\ }\textbf {\bibinfo {volume} {62}},\ \bibinfo
  {pages} {112062} (\bibinfo {year} {2019})}\BibitemShut {NoStop}%
\bibitem [{\citenamefont {Ren}\ \emph {et~al.}(2020{\natexlab{c}})\citenamefont
  {Ren}, \citenamefont {Zhao}, \citenamefont {Zhang},\ and\ \citenamefont
  {Meng}}]{Ren2020NPA}%
  \BibitemOpen
  \bibfield  {author} {\bibinfo {author} {\bibfnamefont {Z.~X.}\ \bibnamefont
  {Ren}}, \bibinfo {author} {\bibfnamefont {P.~W.}\ \bibnamefont {Zhao}},
  \bibinfo {author} {\bibfnamefont {S.~Q.}\ \bibnamefont {Zhang}}, \ and\
  \bibinfo {author} {\bibfnamefont {J.}~\bibnamefont {Meng}},\ }\href {\doibase
  https://doi.org/10.1016/j.nuclphysa.2020.121696} {\bibfield  {journal}
  {\bibinfo  {journal} {Nucl. Phys. A}\ }\textbf {\bibinfo {volume} {996}},\
  \bibinfo {pages} {121696} (\bibinfo {year} {2020}{\natexlab{c}})}\BibitemShut
  {NoStop}%
\bibitem [{\citenamefont {Li}\ \emph {et~al.}(2020)\citenamefont {Li},
  \citenamefont {Ren},\ and\ \citenamefont {Zhao}}]{Li2020PRC}%
  \BibitemOpen
  \bibfield  {author} {\bibinfo {author} {\bibfnamefont {B.}~\bibnamefont
  {Li}}, \bibinfo {author} {\bibfnamefont {Z.~X.}\ \bibnamefont {Ren}}, \ and\
  \bibinfo {author} {\bibfnamefont {P.~W.}\ \bibnamefont {Zhao}},\ }\href
  {\doibase 10.1103/PhysRevC.102.044307} {\bibfield  {journal} {\bibinfo
  {journal} {Phys. Rev. C}\ }\textbf {\bibinfo {volume} {102}},\ \bibinfo
  {pages} {044307} (\bibinfo {year} {2020})}\BibitemShut {NoStop}%
\bibitem [{\citenamefont {Xu}\ \emph {et~al.}(2024{\natexlab{a}})\citenamefont
  {Xu}, \citenamefont {Li}, \citenamefont {Ren},\ and\ \citenamefont
  {Zhao}}]{Xu2024PRC}%
  \BibitemOpen
  \bibfield  {author} {\bibinfo {author} {\bibfnamefont {F.~F.}\ \bibnamefont
  {Xu}}, \bibinfo {author} {\bibfnamefont {B.}~\bibnamefont {Li}}, \bibinfo
  {author} {\bibfnamefont {Z.~X.}\ \bibnamefont {Ren}}, \ and\ \bibinfo
  {author} {\bibfnamefont {P.~W.}\ \bibnamefont {Zhao}},\ }\href {\doibase
  10.1103/PhysRevC.109.014311} {\bibfield  {journal} {\bibinfo  {journal}
  {Phys. Rev. C}\ }\textbf {\bibinfo {volume} {109}},\ \bibinfo {pages}
  {014311} (\bibinfo {year} {2024}{\natexlab{a}})}\BibitemShut {NoStop}%
\bibitem [{\citenamefont {Xu}\ \emph {et~al.}(2024{\natexlab{b}})\citenamefont
  {Xu}, \citenamefont {Li}, \citenamefont {Ring},\ and\ \citenamefont
  {Zhao}}]{Xu2024PLB}%
  \BibitemOpen
  \bibfield  {author} {\bibinfo {author} {\bibfnamefont {F.~F.}\ \bibnamefont
  {Xu}}, \bibinfo {author} {\bibfnamefont {B.}~\bibnamefont {Li}}, \bibinfo
  {author} {\bibfnamefont {P.}~\bibnamefont {Ring}}, \ and\ \bibinfo {author}
  {\bibfnamefont {P.~W.}\ \bibnamefont {Zhao}},\ }\href {\doibase
  https://doi.org/10.1016/j.physletb.2024.138893} {\bibfield  {journal}
  {\bibinfo  {journal} {Phys. Lett. B}\ }\textbf {\bibinfo {volume} {856}},\
  \bibinfo {pages} {138893} (\bibinfo {year} {2024}{\natexlab{b}})}\BibitemShut
  {NoStop}%
\bibitem [{\citenamefont {Xu}\ \emph {et~al.}(2024{\natexlab{c}})\citenamefont
  {Xu}, \citenamefont {Wang}, \citenamefont {Wang}, \citenamefont {Ring},\ and\
  \citenamefont {Zhao}}]{Xu2024PRL}%
  \BibitemOpen
  \bibfield  {author} {\bibinfo {author} {\bibfnamefont {F.~F.}\ \bibnamefont
  {Xu}}, \bibinfo {author} {\bibfnamefont {Y.~K.}\ \bibnamefont {Wang}},
  \bibinfo {author} {\bibfnamefont {Y.~P.}\ \bibnamefont {Wang}}, \bibinfo
  {author} {\bibfnamefont {P.}~\bibnamefont {Ring}}, \ and\ \bibinfo {author}
  {\bibfnamefont {P.~W.}\ \bibnamefont {Zhao}},\ }\href {\doibase
  10.1103/PhysRevLett.133.022501} {\bibfield  {journal} {\bibinfo  {journal}
  {Phys. Rev. Lett.}\ }\textbf {\bibinfo {volume} {133}},\ \bibinfo {pages}
  {022501} (\bibinfo {year} {2024}{\natexlab{c}})}\BibitemShut {NoStop}%
\bibitem [{\citenamefont {Zhao}\ \emph {et~al.}(2010)\citenamefont {Zhao},
  \citenamefont {Li}, \citenamefont {Yao},\ and\ \citenamefont
  {Meng}}]{Zhao2010PRC}%
  \BibitemOpen
  \bibfield  {author} {\bibinfo {author} {\bibfnamefont {P.~W.}\ \bibnamefont
  {Zhao}}, \bibinfo {author} {\bibfnamefont {Z.~P.}\ \bibnamefont {Li}},
  \bibinfo {author} {\bibfnamefont {J.~M.}\ \bibnamefont {Yao}}, \ and\
  \bibinfo {author} {\bibfnamefont {J.}~\bibnamefont {Meng}},\ }\href {\doibase
  10.1103/PhysRevC.82.054319} {\bibfield  {journal} {\bibinfo  {journal} {Phys.
  Rev. C}\ }\textbf {\bibinfo {volume} {82}},\ \bibinfo {pages} {054319}
  (\bibinfo {year} {2010})}\BibitemShut {NoStop}%
\bibitem [{\citenamefont {Simenel}\ and\ \citenamefont
  {Umar}(2014)}]{Simenel2014PRC}%
  \BibitemOpen
  \bibfield  {author} {\bibinfo {author} {\bibfnamefont {C.}~\bibnamefont
  {Simenel}}\ and\ \bibinfo {author} {\bibfnamefont {A.~S.}\ \bibnamefont
  {Umar}},\ }\href {\doibase 10.1103/PhysRevC.89.031601} {\bibfield  {journal}
  {\bibinfo  {journal} {Phys. Rev. C}\ }\textbf {\bibinfo {volume} {89}},\
  \bibinfo {pages} {031601} (\bibinfo {year} {2014})}\BibitemShut {NoStop}%
\bibitem [{\citenamefont {Schunck}\ and\ \citenamefont
  {Robledo}(2016)}]{Schunck2016PPNP}%
  \BibitemOpen
  \bibfield  {author} {\bibinfo {author} {\bibfnamefont {N.}~\bibnamefont
  {Schunck}}\ and\ \bibinfo {author} {\bibfnamefont {L.~M.}\ \bibnamefont
  {Robledo}},\ }\href {\doibase 10.1088/0034-4885/79/11/116301} {\bibfield
  {journal} {\bibinfo  {journal} {Rep. Prog. Phys.}\ }\textbf {\bibinfo
  {volume} {79}},\ \bibinfo {pages} {116301} (\bibinfo {year}
  {2016})}\BibitemShut {NoStop}%
\bibitem [{\citenamefont {Hara}\ and\ \citenamefont
  {Sun}(1995)}]{KENJI1995IJMPE}%
  \BibitemOpen
  \bibfield  {author} {\bibinfo {author} {\bibfnamefont {K.}~\bibnamefont
  {Hara}}\ and\ \bibinfo {author} {\bibfnamefont {Y.}~\bibnamefont {Sun}},\
  }\href {\doibase 10.1142/S0218301395000250} {\bibfield  {journal} {\bibinfo
  {journal} {Int. J. Mod. Phys. E}\ }\textbf {\bibinfo {volume} {04}},\
  \bibinfo {pages} {637} (\bibinfo {year} {1995})}\BibitemShut {NoStop}%
\bibitem [{\citenamefont {Ma}\ and\ \citenamefont {Ma}(2018)}]{Ma2018PPNP}%
  \BibitemOpen
  \bibfield  {author} {\bibinfo {author} {\bibfnamefont {C.-W.}\ \bibnamefont
  {Ma}}\ and\ \bibinfo {author} {\bibfnamefont {Y.-G.}\ \bibnamefont {Ma}},\
  }\href {\doibase https://doi.org/10.1016/j.ppnp.2018.01.002} {\bibfield
  {journal} {\bibinfo  {journal} {Progress in Particle and Nuclear Physics}\
  }\textbf {\bibinfo {volume} {99}},\ \bibinfo {pages} {120} (\bibinfo {year}
  {2018})}\BibitemShut {NoStop}%
\bibitem [{\citenamefont {Li}\ \emph {et~al.}(2023{\natexlab{b}})\citenamefont
  {Li}, \citenamefont {Vretenar}, \citenamefont {Nik\ifmmode \check{s}\else
  \v{s}\fi{}i\ifmmode~\acute{c}\else \'{c}\fi{}}, \citenamefont {Zhao},\ and\
  \citenamefont {Meng}}]{Li2023PRC_gdTDGCM}%
  \BibitemOpen
  \bibfield  {author} {\bibinfo {author} {\bibfnamefont {B.}~\bibnamefont
  {Li}}, \bibinfo {author} {\bibfnamefont {D.}~\bibnamefont {Vretenar}},
  \bibinfo {author} {\bibfnamefont {T.}~\bibnamefont {Nik\ifmmode
  \check{s}\else \v{s}\fi{}i\ifmmode~\acute{c}\else \'{c}\fi{}}}, \bibinfo
  {author} {\bibfnamefont {P.~W.}\ \bibnamefont {Zhao}}, \ and\ \bibinfo
  {author} {\bibfnamefont {J.}~\bibnamefont {Meng}},\ }\href {\doibase
  10.1103/PhysRevC.108.014321} {\bibfield  {journal} {\bibinfo  {journal}
  {Phys. Rev. C}\ }\textbf {\bibinfo {volume} {108}},\ \bibinfo {pages}
  {014321} (\bibinfo {year} {2023}{\natexlab{b}})}\BibitemShut {NoStop}%
\bibitem [{\citenamefont {Li}\ \emph {et~al.}(2024{\natexlab{d}})\citenamefont
  {Li}, \citenamefont {Vretenar}, \citenamefont {Nik\ifmmode \check{s}\else
  \v{s}\fi{}i\ifmmode~\acute{c}\else \'{c}\fi{}}, \citenamefont {Zhao},
  \citenamefont {Zhao},\ and\ \citenamefont {Meng}}]{Li2024Fop}%
  \BibitemOpen
  \bibfield  {author} {\bibinfo {author} {\bibfnamefont {B.}~\bibnamefont
  {Li}}, \bibinfo {author} {\bibfnamefont {D.}~\bibnamefont {Vretenar}},
  \bibinfo {author} {\bibfnamefont {T.}~\bibnamefont {Nik\ifmmode
  \check{s}\else \v{s}\fi{}i\ifmmode~\acute{c}\else \'{c}\fi{}}}, \bibinfo
  {author} {\bibfnamefont {J.}~\bibnamefont {Zhao}}, \bibinfo {author}
  {\bibfnamefont {P.~W.}\ \bibnamefont {Zhao}}, \ and\ \bibinfo {author}
  {\bibfnamefont {J.}~\bibnamefont {Meng}},\ }\href {\doibase
  10.1007/s11467-023-1381-4} {\bibfield  {journal} {\bibinfo  {journal}
  {Frontiers of Physics}\ }\textbf {\bibinfo {volume} {19}},\ \bibinfo {pages}
  {44201} (\bibinfo {year} {2024}{\natexlab{d}})}\BibitemShut {NoStop}%
\bibitem [{\citenamefont {Li}\ \emph {et~al.}(2025)\citenamefont {Li},
  \citenamefont {Vretenar}, \citenamefont {Nik\ifmmode \check{s}\else
  \v{s}\fi{}i\ifmmode~\acute{c}\else \'{c}\fi{}}, \citenamefont {Zhao},\ and\
  \citenamefont {Meng}}]{Li2025PRC}%
  \BibitemOpen
  \bibfield  {author} {\bibinfo {author} {\bibfnamefont {B.}~\bibnamefont
  {Li}}, \bibinfo {author} {\bibfnamefont {D.}~\bibnamefont {Vretenar}},
  \bibinfo {author} {\bibfnamefont {T.}~\bibnamefont {Nik\ifmmode
  \check{s}\else \v{s}\fi{}i\ifmmode~\acute{c}\else \'{c}\fi{}}}, \bibinfo
  {author} {\bibfnamefont {P.~W.}\ \bibnamefont {Zhao}}, \ and\ \bibinfo
  {author} {\bibfnamefont {J.}~\bibnamefont {Meng}},\ }\href {\doibase
  10.1103/PhysRevC.111.L051302} {\bibfield  {journal} {\bibinfo  {journal}
  {Phys. Rev. C}\ }\textbf {\bibinfo {volume} {111}},\ \bibinfo {pages}
  {L051302} (\bibinfo {year} {2025})}\BibitemShut {NoStop}%
\bibitem [{\citenamefont {Hu}\ \emph {et~al.}(2014)\citenamefont {Hu},
  \citenamefont {Gao},\ and\ \citenamefont {Chen}}]{Hu2014PLB}%
  \BibitemOpen
  \bibfield  {author} {\bibinfo {author} {\bibfnamefont {Q.~L.}\ \bibnamefont
  {Hu}}, \bibinfo {author} {\bibfnamefont {Z.~C.}\ \bibnamefont {Gao}}, \ and\
  \bibinfo {author} {\bibfnamefont {Y.~S.}\ \bibnamefont {Chen}},\ }\href
  {\doibase https://doi.org/10.1016/j.physletb.2014.05.045} {\bibfield
  {journal} {\bibinfo  {journal} {Phys. Lett. B}\ }\textbf {\bibinfo {volume}
  {734}},\ \bibinfo {pages} {162} (\bibinfo {year} {2014})}\BibitemShut
  {NoStop}%
\end{thebibliography}

%\end{CJK*}
\end{document}